%% file: main.tex
\documentclass[usenatbib,usegraphicx]{mn2e}
\usepackage{amssymb}
\usepackage[fleqn]{amsmath}
\usepackage{charter}
\usepackage{mathdesign}
\usepackage{url}
\usepackage[hyperfootnotes, hyperfigures,bookmarks, pdftex]{hyperref}
\hypersetup{bookmarksnumbered, bookmarksopen, colorlinks, citecolor=blue, pdfduplex=DuplexFlipLongEdge}
\hypersetup{pdftitle=Mesoscale disc Turbulence, pdfauthor=Jacob B. Simon, pdfsubject=Astrophysics}
\usepackage{bookmark}
\providecommand{\adsurl}[1]{\href{#1}{ADS}}

\pagerange{\pageref{firstpage}--\pageref{lastpage}} \pubyear{2004}

\def\LaTeX{L\kern-.36em\raise.3ex\hbox{a}\kern-.15em
T\kern-.1667em\lower.7ex\hbox{E}\kern-.125emX}

\newcommand{\bm}[1]{\mbox{\boldmath{$#1$}}}

\newcommand{\del}{{\bf \nabla}}
\newcommand{\alf}{{\rm Alfv\acute{e}n}}

\newcommand{\cs}{c_{\rm s}}  
\newcommand{\va}{v_{\rm A}}

\newcommand{\pv}{P_{\rm var}}

\newcommand{\half}{{\rm FT0.5}}
\newcommand{\two}{{\rm FT2}}
\newcommand{\four}{{\rm FT4}}
\newcommand{\eight}{{\rm FT8}}

\newcommand{\sixteeny}{{\rm Y16}}

\input shortcuts
\title[Accretion Disc Turbulence]{Emergent Mesoscale Phenomena in Magnetized Accretion Disc Turbulence}

\author[Simon et al.]
{Jacob B. Simon$^{1 \ast}$,
Kris Beckwith$^{1,3 \star}$,
Philip J. Armitage$^{1,2 \dagger}$
\\ $^1$JILA,
University of Colorado \& NIST,
440 UCB,
Boulder, CO 80309-0440
\\ $^2$Department of Astrophysical and Planetary Sciences, 
University of Colorado at Boulder
\\ $^3$Tech-X Corporation,
5621 Arapahoe Ave. Suite A
Boulder, CO 80303
\\Email: $^\ast$jbsimon@jila.colorado.edu,
$^\star$kris.beckwith@jila.colorado.edu,
$^\dagger$pja@jilau1.colorado.edu}

\begin{document}

\label{firstpage}

\maketitle

\begin{abstract} 
We study how the structure and variability of magnetohydrodynamic (MHD) turbulence in accretion discs converge with domain size.
Our results are based on a series of vertically stratified local 
simulations, computed using the {\em Athena} MHD code, that have fixed spatial resolution, but 
varying radial and azimuthal extent (from $\Delta R = 0.5 H$ to $16 H$, where $H$ is the vertical 
scale height). We show that elementary local diagnostics of the turbulence, including the Shakura-Sunyaev 
$\alpha$ parameter, the ratio of Maxwell stress to magnetic energy, and the ratio of magnetic to 
fluid stresses, converge to within the precision of our measurements  
for spatial domains of radial size $L_x \ge 2 H$. We obtain $\alpha \simeq 0.02 - 0.03$, 
consistent with other recent determinations. Very small domains ($L_x = 0.5 H$) return 
anomalous results, independent of spatial resolution. This convergence with domain size, however, is only valid for a limited 
set of diagnostics: larger 
spatial domains admit the emergence of dynamically important mesoscale structures. In 
our largest simulations, the Maxwell stress shows a significant large scale non-local 
component, while the density develops long-lived axisymmetric perturbations (``zonal flows") 
at the 20\% level. Most strikingly, the variability of the disc in fixed-sized patches decreases 
strongly as the simulation volume increases, while variability in the magnetically dominated 
corona remains constant. Comparing our largest local simulations to global simulations with 
comparable spatial resolution, we find generally good agreement. There is no direct 
evidence that the presence of curvature terms or radial gradients in global 
calculations materially affect the turbulence, except to perhaps introduce an outer radial scale for mesoscale structures. The demonstrated importance of 
mean magnetic fields -- seen in both large local and global simulations -- implies, however,  
that the growth and saturation of these fields is likely of critical importance for the evolution of accretion discs.
\end{abstract}


\input intro

\input numerics

\newpage
\input results

\input comparison

\input conclusion


\section*{Acknowledgements}
We thank Mitch Begelman, Charles Gammie, Julian Krolik, Omer Blaes, Shigenobu Hirose, Kareem Sorathia, Jim Stone, Xioayue Guan, Shane Davis, and John Hawley for useful discussions and advice. We also thank an anonymous referee for useful comments on an earlier draft of this paper. 
This work was supported by the NSF under grant numbers AST-0807471 and AST-0907872, and by NASA under grant numbers NNX09AB90G and NNX11AE12G. This research was supported in part by the NSF through TeraGrid resources provided by the Texas Advanced Computing Center and the National Institute for Computational Science under grant number TG-AST090106.  We also acknowledge the Texas Advanced Computing Center at The University of Texas at Austin for providing HPC and visualization resources that have contributed to the research results reported within this paper. Computations were also performed on Kraken at the National Institute for Computational Sciences.


\bibliography{references}

\label{lastpage}

\end{document}

%% file: intro.tex
\section{Introduction}

Accretion of gas through a disc onto a central object 
plays a pivotal role in the formation of protostars and planets, and furnishes the only probe of the 
environment around black holes to date.  If the disc is sufficiently ionized, accretion is driven by a powerful instability that is present in orbiting, magnetized gas: the magnetorotational instability \cite[MRI;][]{balbus:1998,Balbus:1991,Blaes:1994,Goodman:1994}.  The MRI gives rise to sustained
magnetohydrodynamic (MHD) turbulence in which the turbulent fluctuations are correlated
to produce a positive stress tensor and thus the outward angular momentum transport necessary for the gas to accrete. 
While numerous analytic investigations of the MRI's linear regime \cite[e.g.,][]{Balbus:1991,Blaes:1994} have been insightful, a complete understanding of turbulent disc accretion requires the ability to probe the fully nonlinear behavior of this instability.  This is best done through the use of numerical simulations \cite[though, some
analytic work in the non-linear regime has been done;][]{Goodman:1994,Pessah:2009}.

Numerical studies of accretion discs can be either global or local. 
These classes of simulations are distinguished by the different length scales that are captured. A local patch of accretion disc can be characterized by two length scales: the distance from the central object, $R_0$ and the local scale height (disc thickness), $H \approx c_s/ \Omega$ (where $c_s$ and $\Omega$ are the sound speed and angular velocity at $R_0$, respectively). If we associate the domain of a simulation with a length scale, $L$, then local simulations are characterized by $H \sim L \ll R_0$, whereas global simulations are characterized by $H \ll L \sim R_0$.

Local simulations invariably employ the shearing box approximation \cite[e.g.,][]{Hawley:1995,Brandenburg:1995}, in which the MHD equations are 
solved in the frame of a local, co-rotating patch of the disc. The assumption that $H \sim L \ll R_0$ allows the use of a Cartesian geometry while retaining the essential dynamics of differential rotation. This simplifies the problem down to its basic ingredients, orbital shear and magnetization, allowing one to answer basic questions
about MRI-driven turbulence, such as ``what determines the amplitude of turbulent fluctuations?"
At fixed computational cost, local simulations provide the best resolution of small-scale features in the turbulence, 
which is particularly advantageous if one wishes to study processes such as resistivity or viscosity that are physically 
important only for $l \ll H$. Local simulations can in principle be simplified yet further by ignoring the vertical 
component of gravity (``unstratified" simulations). However, in our current work, we only consider stratified calculations which 
include the vertical gravity from the central object.

Global simulations \cite[e.g.,][]{Armitage:1998,hawley:2000}, on the other hand, evolve a domain that is comparable to the distance from the central object.  These simulations are not only typically larger in spatial extent than local models -- which may be important if for example long wavelength radial and azimuthal fluctuations are present in the disc -- they also include new physical effects. In particular, global models include terms associated with disc curvature, and possess the asymmetry between inward and outward radial directions that is needed to explicitly represent mass accretion. However, these advantages come at the expense of non-trivial ``inner" and ``outer" boundary conditions.  For example, with the exception of black hole disk simulations, where an outflow boundary is applied at the inner boundary to account for material plummeting into an event horizon, particular methods (e.g., viscous or magnetic damping) have to be employed to reduce spurious effects from these boundaries.  Global simulations also suffer from poor resolution of small scales, such that most global simulations to-date fail to meet the minimum requirements for numerical convergence that have been derived from local models \citep{Hawley:2011}. Only recently have increases in computational power and algorithmic improvements made it possible to compute global simulations at the resolutions required to adequately resolve the MRI at the level done in local simulations \citep{Noble:2010,Beckwith:2011,Sorathia:2011}.

The near congruence of resolutions between local and global simulations makes possible a comparison of the turbulent properties in these two regimes. One question is whether a local patch of a well-resolved global simulation resembles an equivalently resolved local simulation? If not, are any discrepancies due solely to the different sizes of the domains, or are they instead attributable to truly global effects such as curvature? One approach to addressing this question is to inspect the properties of local sub-domains within a global simulation and compare the statistical properties of these sub-domains to results from shearing boxes \citep{Sorathia:2010,Sorathia:2011}. These studies suggest that some of the basic properties of local MRI turbulence, such as the relationship between turbulent stress and magnetic flux through a local patch \citep{Hawley:1995} and the tilt angle of the magnetic field autocorrelation function \citep{Guan:2009a}, carry over to global calculations.

A complementary approach is to examine the properties of turbulence in large ``mesoscale" simulations, characterized by $H \ll L \ll R_0$ \citep{Guan:2011}. These simulations, which continue to utilize the shearing box approximation and are in this sense still local, allow study of the influence of long wavelength azimuthal and radial modes on the properties of the turbulence, while also capturing small scale turbulent fluctuations. Compared to global simulations, the mesoscale neglects the effect of curvature terms on the disc, which, in principle, allows the influence of these terms to be elucidated. \cite{Guan:2011} \cite[see also][]{Davis:2010} find that, in the mesoscale regime, turbulence is characterized by small scale fluctuations regardless of domain size. These authors also find that, away from the mid-plane, in the coronal region, the magnetic field is correlated on scales of $\sim 10H$ but does not contribute significantly to angular momentum transport. 
Larger scales are not, however, unimportant. 
\cite{Nelson:2010} \cite[see also][]{Gressel:2011,Yang:2011} demonstrate that long wavelength (again $\gg H$) density correlations determine the amplitude of stochastic density fluctuations (spiral density waves), while the calculations of \cite{Johansen:2009} show long lived density/pressure zonal flows in geostrophic balance.

Our goal in this work is to systematically study how the properties of disc turbulence change as one transitions 
from the local to the mesoscale regime. We do so though a series of shearing box simulations of increasing domain size, from $L_x = 0.5H\times L_y = 2H\times L_z = 8H$ to $L_x = 16H\times L_y = 32H\times L_z = 8H$ (here $x$, $y$ and $z$ refer to the radial, azimuthal, and vertical directions respectively)\footnote{Whether the neglect of radial gradients and curvature terms in our large local boxes is physically justified depends upon the geometric thickness of the disc being modeled: a domain of radial scale $\gg H$ remains ``local" in thin regions of an AGN disc where $H / R \leq 10^{-2}$, whereas the same would not be true of a thicker protoplanetary disc. Our primary interest in the large boxes is as well-controlled model systems, though the actual value of $H/R$ should be borne in mind when comparing our results to global simulations.}. In particular, we wish to answer the following questions.  How do various properties of MRI-driven turbulence depend on box size? As we push towards the mesoscale regime, what if any phenomena emerge at these larger scales?  Is there a scale at which the turbulent properties resemble those in a global disc calculation?

The rest of this paper is structured as follows. In \S\ref{Method}, we give details of the numerical algorithm used to integrate the equations of ideal MHD in the local limit and the initial and boundary conditions used for the simulations. In \S\ref{steadystate}, we detail the properties of the quasi-stationary turbulent state via volume-averaged quantities and the vertical structure of the turbulence. In \S\ref{mesoscale}, we examine the properties of mesoscale structures within the magnetic field and the implications for the locality of angular momentum transport. The derived variability of accretion is examined in \S\ref{var}. In \S\ref{compare}  we compare a subset of our results to those obtained from well-resolved global simulations. Finally, in \S\ref{discussion}, we summarize our results and discuss implications for the physics of magnetized accretion discs.

%% file: numerics.tex
\section{Method}
\label{Method}

\subsection{Numerical Algorithm}
\label{numerics}

For these calculations, we use {\it Athena}, a second-order accurate Godunov
flux-conservative code for solving the equations of MHD.  \textit{Athena}
uses the dimensionally unsplit corner transport upwind (CTU) method
of \cite{colella:1990} coupled with the third-order in space piecewise
parabolic method (PPM) of \cite{Colella:1984} and a constrained transport
\citep[CT;][]{evans:1988} algorithm for preserving the $\del \cdot {\bm
B}$~=~0 constraint.  We use the HLLD Riemann solver to calculate the
numerical fluxes \cite[]{miyoshi:2005,Mignone:2007b}.  A detailed description
of the \textit{Athena} algorithm and the results of various test problems
are given in \cite{Gardiner:2005}, \cite{Gardiner:2008}, and \cite{Stone:2008}.

The simulations use the shearing box approximation, a model
for a local co-rotating disc patch whose size is small compared to the
radial distance from the central object, $R_0$.  We construct a local
Cartesian frame, $x=(R-R_0)$, $y=R_0 \phi$, and $z$,
co-rotating with an angular velocity $\Omega$ corresponding to
the orbital frequency at $R_0$, the center of the box.  In this 
frame, the equations of motion become \citep{Hawley:1995}:

\begin{equation}
\label{sbeqns}
\begin{split}
\frac{\partial \rho}{\partial t} + \del \cdot (\rho {\bm v}) = 0 \\
\frac{\partial \rho {\bm v}}{\partial t} + \del \cdot \left(\rho {\bm v}{\bm v} - {\bm B}{\bm B}\right) + \del \left(P + \frac{1}{2} B^2\right) \\
= 2 q \rho \Omega^2 {\bm x} - \rho \Omega^2 {\bm z} - 2 {\bm \Omega} \times \rho {\bm v} \\
\frac{\partial {\bm B}}{\partial t} - \del \times \left({\bm v} \times {\bm B}\right) = 0
\end{split}
\end{equation} 

\noindent 
where $\rho$ is the mass density, $\rho {\bm v}$ is the momentum
density, ${\bm B}$ is the magnetic field, $P$ is the gas pressure,
and $q$ is the shear parameter, defined as $q = -d$ln$\Omega/d$ln$R$.
We use $q = 3/2$, appropriate for a Keplerian disc.  We
assume an isothermal equation of state $P = \rho \cs^2$, where $\cs$
is the isothermal sound speed.  From left to right, the source terms
in the momentum equation correspond to radial tidal forces
(gravity and centrifugal), vertical gravity, and the Coriolis force.
Note that our system of units has the magnetic permeability $\mu = 1$.

The numerical integration of the shearing box equations require additions to the \textit{Athena} algorithm, the 
details of which can be found in \cite{Stone:2010} and in the Appendix of \cite{Simon:2011a}.  Briefly, we utilize
Crank-Nicholson differencing to conserve epicyclic motion exactly and orbital advection to subtract 
off the background shear flow \cite[]{Stone:2010}.  The $y$ boundary conditions are strictly periodic, whereas
the $x$ boundaries are shearing periodic \cite[]{Hawley:1995,Simon:2011a}. The vertical boundaries are the
outflow boundary conditions described in \cite{Simon:2011a}.

\begin{table}
\begin{center}
\caption{Shearing Box Simulations\label{tbl:sims}}
\begin{tabular}{ @{}l|cccccccc}
\hline
Label &
Domain Size&
Resolution &
Initial Field \\
 & $(L_x/H \times L_y/H \times L_z/H) $ &
(zones$/H$) &
Configuration \\
\hline
\hline
\half & $0.5 \times 2 \times 8$ & 32 & Flux tube\\
FT0.5m & $0.5 \times 2 \times 8$ & 72 & Flux tube \\
FT0.5h & $0.5 \times 2 \times 8$ & 144 & Flux tube \\
\two & $2 \times 4 \times 8$ & 32 & Flux tube\\
\four & $4 \times 8 \times 8$ & 32 & Flux tube\\
\eight & $8 \times 16 \times 8$ & 32 & Flux tube\\
\sixteeny &  $16 \times 32 \times 8$ & 36 & Toroidal\\
\hline
\end{tabular}
\end{center}

\medskip Here, the scale height is defined as $H = \sqrt{2}\cs/\Omega$.

\end{table}

\subsection{Simulation Parameters and Initial Conditions}
\label{parameters}

All of our simulations are vertically stratified, with an initial density
corresponding to isothermal hydrostatic equilibrium.

\begin{equation}
\label{density_init}
\rho(x,y,z) = \rho_0 {\rm exp}\left(-\frac{z^2}{H^2}\right),
\end{equation}

\noindent where $\rho_0 = 1$ is the mid-plane density, and $H$
is the scale height in the disc,

\begin{equation}
\label{scale_height}
H = \frac{\sqrt{2} \cs}{\Omega}.
\end{equation}

\noindent The isothermal sound speed, $\cs = 7.07 \times 10^{-4}$,
corresponding to an initial value for the mid-plane gas pressure of $P_0 = 5
\times 10^{-7}$.  With $\Omega = 0.001$, the value for the scale height
is $H = 1$.   A density floor of $10^{-4}$ ($10^{-6}$ for the smallest domain run) is applied to the physical domain as
too small a density leads to a large $\alf$ speed and a very small
timestep.  Furthermore, numerical errors make it difficult
to evolve regions of very small plasma $\beta$.  

 \begin{figure}
\begin{center}
\leavevmode
\includegraphics[width=0.5\textwidth]{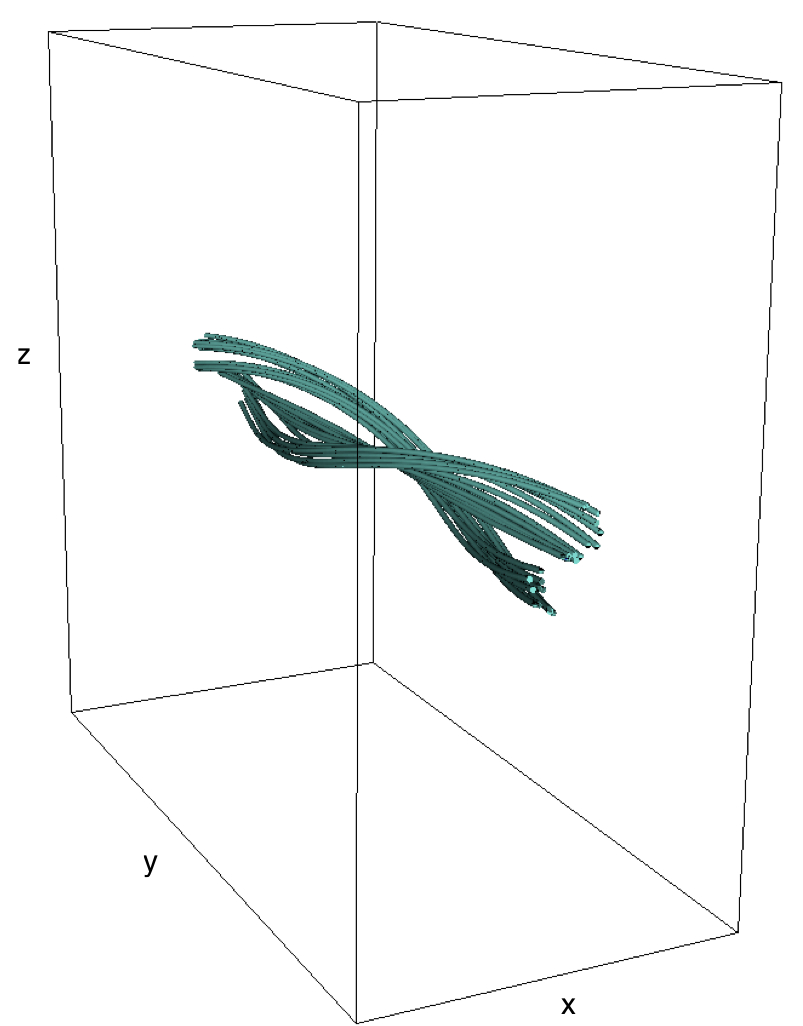}
\end{center}
\caption[]{Rendering of the initial magnetic field lines for the $\four$ run.  This is the twisted azimuthal flux tube of \cite{Hirose:2006}, and it is the initial field configuration for all of our runs except for $\sixteeny$.}
\label{init_field}
\end{figure}

The initial magnetic field configuration for most of our calculations is the
twisted azimuthal flux tube of \cite{Hirose:2006}, with minor modifications
to the dimensions and $\beta$ values.
In particular, the initial toroidal field, $B_y$, is given by
\begin{equation}
\label{toroidal}
B_y = \left\{ \begin{array}{ll}
\sqrt{\frac{2 P_0}{\beta_y} - \left(B_x^2+B_z^2\right)}& \quad
\mbox{if $B_x^2 + B_z^2 \neq 0$} \\
0 & \quad 
\mbox{if $B_x^2 + B_z^2 = 0$}
\end{array} \right.
\end{equation}
\noindent with $\beta_y = 100$.
The poloidal field components, $B_x$ and $B_z$,
are calculated from the $y$ component of the vector potential,
\begin{equation}
\label{vector_potential}
A_y = \left\{ \begin{array}{ll}
- \sqrt{\frac{2 P_0}{\beta_p}}\frac{R_0}{\pi} \left[1 + {\rm cos}\left(\frac{\pi r}{R_0}\right)\right] & \quad
\mbox{if $r < R_0$} \\
0 & \quad 
\mbox{if $r \geq R_0$}
\end{array} \right.
\end{equation}
\noindent where $r = \sqrt{x^2 + z^2}$ and $\beta_p = 1600$ is the
poloidal field $\beta$ value.  We choose $R_0$ to always be one fourth of the radial domain size; $R_0 = L_x/4$. Figure~\ref{init_field} shows a rendering of the initial magnetic field configuration from $\four$.

 For the largest domain size, the aspect ratio is such that $R_0 = L_z/2$. As was evident from an initial run at this size and with the flux tube field geometry, the flux tube cannot be properly contained within the domain without initial and dominant boundary condition effects.  Thus, we chose our largest domain run, $\sixteeny$, to be initialized with a constant $\beta = 100$ toroidal field throughout the entire domain. For all of our calculations, random perturbations are added to the density and velocity components to seed the MRI. Table~\ref{tbl:sims} lists the simulations that we have analyzed for this work. The label for each calculation describes the initial field geometry and the $x$ domain size.  So, for run $\half$, FT denotes the flux tube geometry, and the 0.5 denotes $L_x = 0.5 H$.   As shown in the table, we have also run two higher resolution simulations equivalent to FT0.5; FT0.5m is carried out at 72 zones per $H$ and FT0.5h is carried out at 144 zones per $H$.

%% file: results.tex
\section{Characterizing the Turbulent Steady State}\label{steadystate}

\begin{figure}
\begin{center}
\leavevmode
\includegraphics[width=\columnwidth]{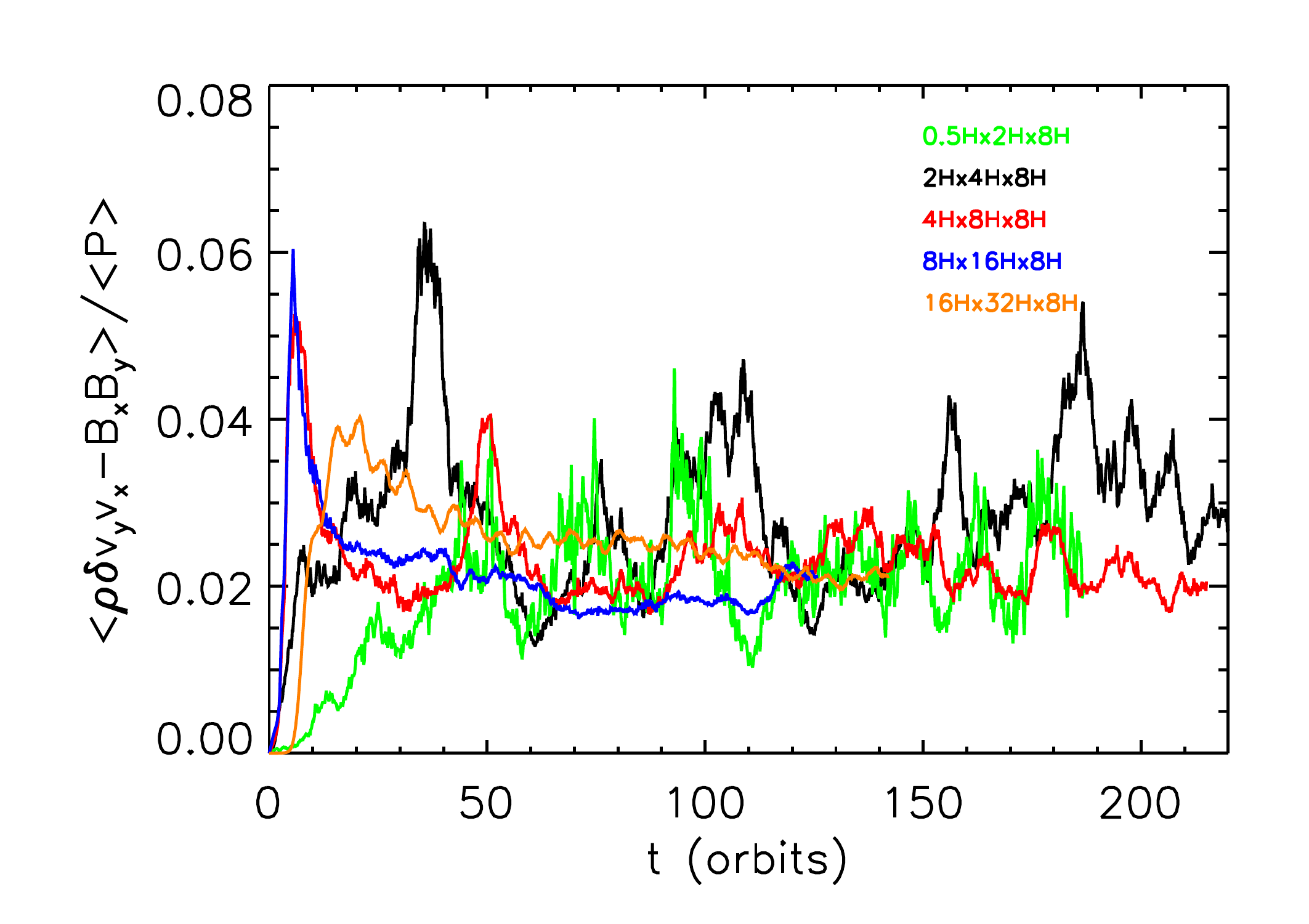}
\end{center}
\caption[]{Time history of volume-averaged Maxwell and Reynolds stresses normalized by the volume-averaged pressure for each shearing box domain size.
The volume average is done for all $x$ and $y$ and for $|z| < 2H$.  The saturation level is roughly consistent between all of the calculations.} 
\label{stress2H}
\end{figure}

Although the focus of the current study is on the convergence of the properties of disc turbulence with varying 
domain size, this can only be meaningful if the simulations being compared have adequate spatial resolution. 
To satisfy ourselves that this is the case, we first compute the quality parameter $Q_j$ \citep{Hawley:2011}, 
which characterizes the effective numerical resolution of the turbulence (i.e., how well-resolved the turbulence is). For a direction $j=y,z$, the quality parameter is defined as, 
\begin{equation}
Q_j = \overline{\frac{2\pi|\va|}{\Omega \Delta x_j}} ; \;\;
|\va| \equiv \frac{\langle B_j^2\rangle}{\langle\rho\rangle}.
\label{quality}
\end{equation}
\noindent where $\Delta x_j$ is the grid cell spacing along direction $j$. Here, the brackets denote a volume average for all $x$ and $y$ and for $|z| < 2H$, and the overbar is a time average from orbit 50 onwards. \cite{Sano:2004} demonstrate that $Q\sim6$ is required in order for the MRI to be well-resolved in the saturated state (though this value is likely to be code-dependent). The values of $Q$ measured for our simulations are given in the final two columns of Table \ref{tbl:sat}. All of the simulations are well-resolved according to this criterion.

Turning to physical rather than numerical quantities, we examine the total (Maxwell plus Reynolds) stress normalized to the gas pressure. The time evolution of this measure, which is equivalent to the $\alpha$-parameter introduced by \cite{Shakura:1973}, is shown in Figure \ref{stress2H}, where we have volume-averaged the data over all $x$ and $y$ and for $|z| < 2H$. Following an initial transient period, which lasts for the first $\sim50$ orbits of the evolution, the normalized stress levels are all roughly equal between different domain sizes.  There is, however, a dramatic change in temporal variability as one goes to larger box sizes.  This effect was present in the simulations of \cite{Davis:2010}, and we will discuss it in detail in \S\ref{var}.

\begin{table}
\begin{center}
\caption{Saturation Characteristics\label{tbl:sat}}
\begin{tabular}{ @{}l|cccccccc}
\hline
Run&
$\alpha$&
$\alpha_{\rm mag}$&
MR&
${\rm BR}_x$ &
${\rm BR}_y$ &
${\rm BR}_z$ &
$Q_y$ &
$Q_z$ \\
\hline
\hline
\half & 0.022 & 0.16 & 4.6 & 0.056 & 0.92 & 0.022 & 67 & 10 \\
FT0.5m & 0.030 & 0.13 & 4.7 & 0.057 & 0.92 & 0.024 & 188 & 30 \\
FT0.5h & 0.032 & 0.22 & 5.5 & 0.096 & 0.86 & 0.045 & 297 & 67 \\
\two & 0.027 & 0.38 & 3.8 & 0.11 & 0.84 & 0.051 & 44 & 11  \\
\four & 0.023 & 0.40 & 3.5 & 0.11 & 0.84 & 0.048 & 39 & 9 \\
\eight & 0.020& 0.41 & 3.4 & 0.11 & 0.85 & 0.045 & 35 & 8 \\
\sixteeny & 0.024 & 0.40 & 3.5 & 0.12 & 0.83 & 0.050 & 44 & 11 \\
\hline
\end{tabular}
\end{center}

\medskip Here, $\alpha$ is the ratio of stress to gas pressure (i.e., the traditional \cite{Shakura:1973} $\alpha$ parameter), $\alpha_{\rm mag}$
is the ratio of Maxwell stress to magnetic energy, MR is the Maxwell to Reynolds stress ratio, ${\rm BR}_i$ is the ratio of magnetic energy component $i$
to the total magnetic energy, and $Q_i$ is the quality parameter along direction $i$ as defined via equation~(\ref{quality}).

\end{table}

\begin{figure}
\begin{center}
\leavevmode
\includegraphics[width=0.45\textwidth]{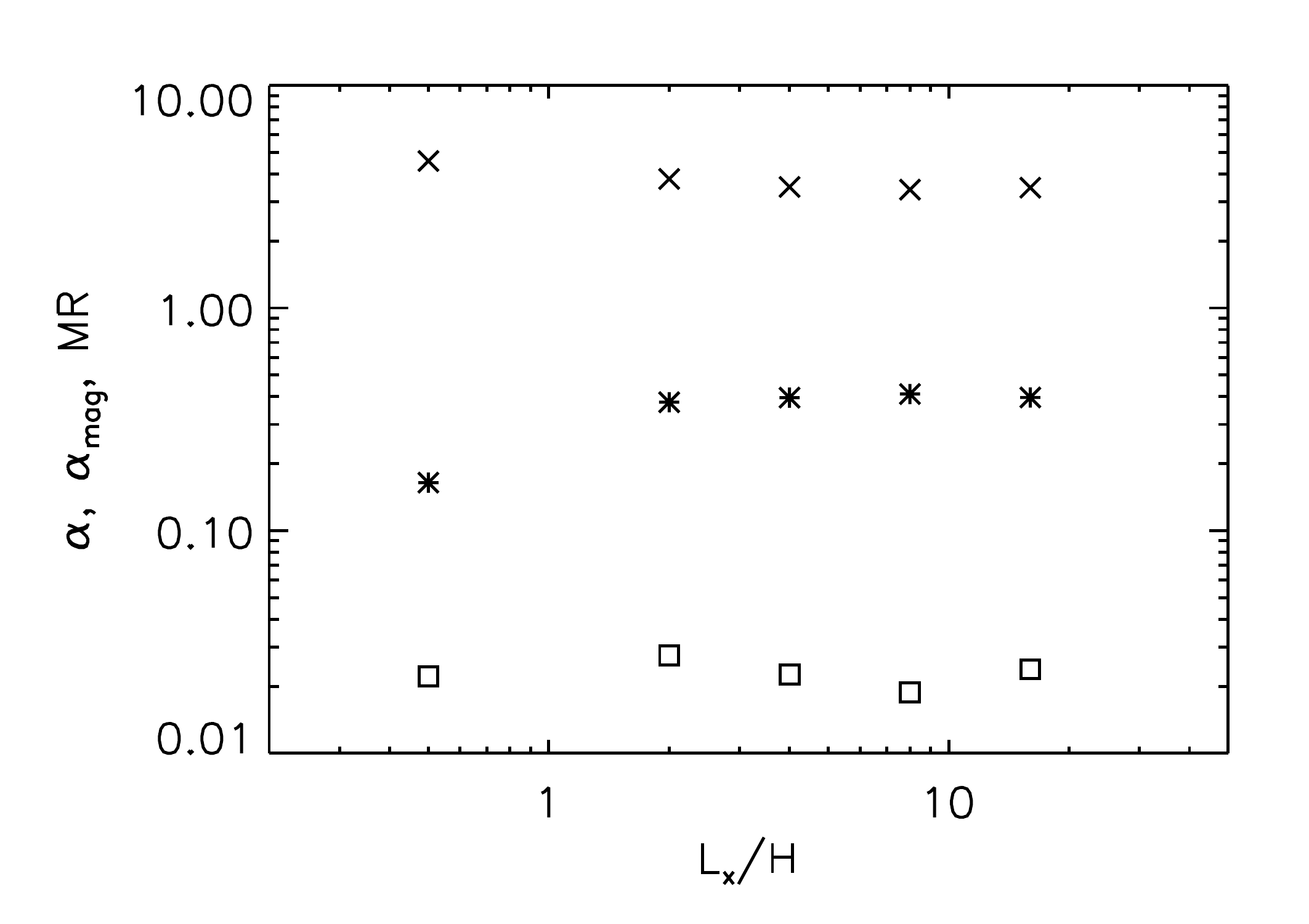}
\includegraphics[width=0.45\textwidth]{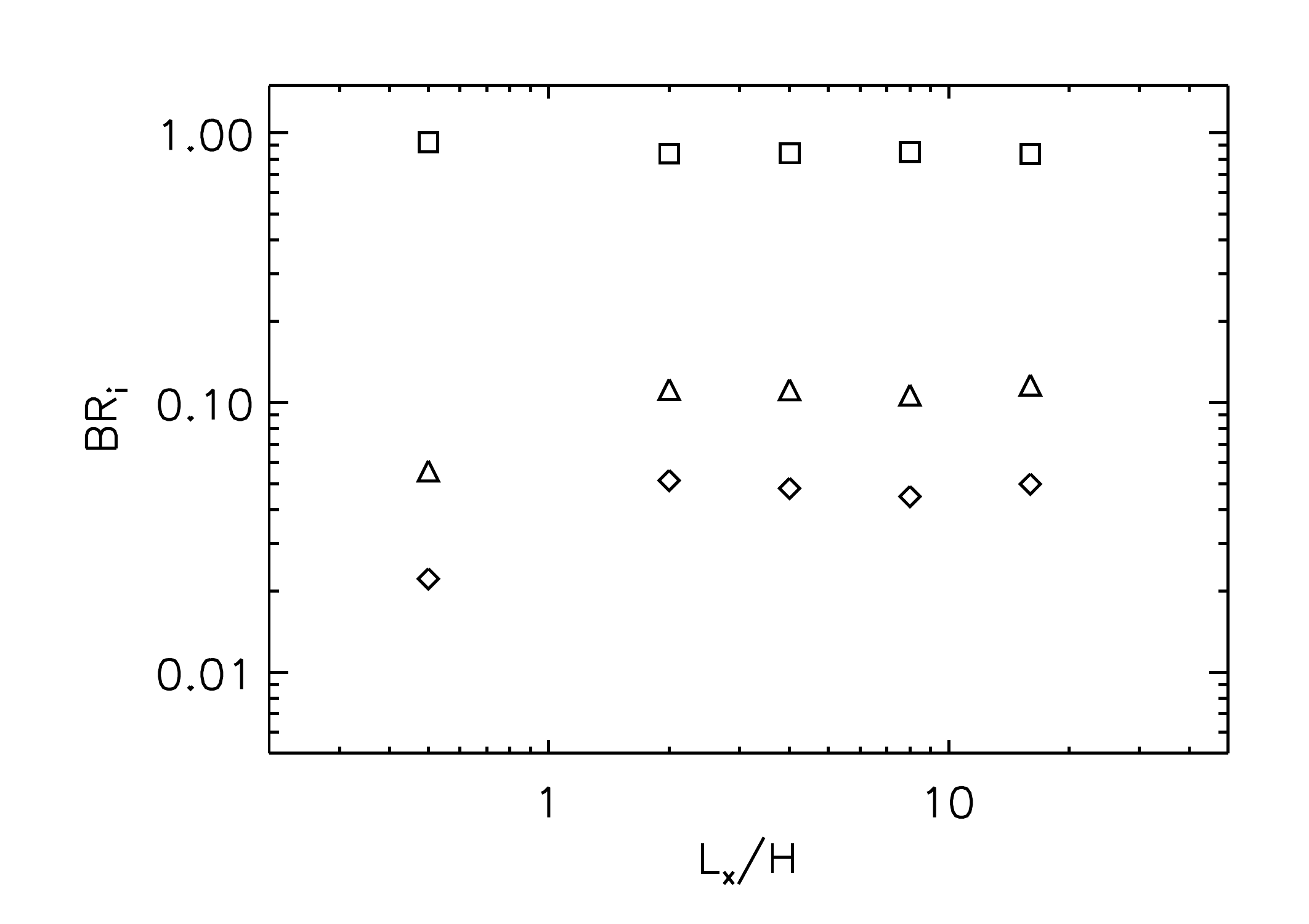}
\end{center}
\caption[]{Several time- and volume-averaged quantities versus radial domain size.  The left panel displays $\alpha$ (squares), $\alpha_{\rm mag}$ (asterisks),
and MR (x's), whereas the right panel displays BR$_y$ (squares), BR$_x$ (triangles), and BR$_z$ (diamonds).  These diagnostics are defined by equation~(\ref{diagnostics}) and converge for $L_x/H \gtrsim 2$.}
\label{turb_avg}
\end{figure}

The data of Figure \ref{stress2H} allow us to define a time-period, after the $50$-orbit mark, where the flow is in a quasi-stationary turbulent state. We use the simulation data from this period to construct various time- and volume-averaged measures of the turbulence (where the volume-average is again performed for all $x$ and $y$ and for $|z| < 2H$). We consider the following diagnostics that characterize the \emph{physical} state of the turbulence:
\begin{equation}
\label{diagnostics}
\begin{split}
\alpha \equiv \overline{\frac{\langle \rho v_x\delta v_y-B_xB_y\rangle}{\langle P\rangle}}
\;\; ; \;\;
\alpha_{\rm mag} \equiv \overline{\frac{2\langle -B_xB_y\rangle}{\langle B^2\rangle}} \\
{\rm MR} \equiv \overline{\frac{\langle-B_xB_y\rangle}{\langle \rho v_x\delta v_y\rangle}} \;\; ; \;\;
{\rm BR}_i \equiv \overline{\frac{\langle B_i^2\rangle}{\langle B^2\rangle}}.
\end{split}
\end{equation}
These are, in turn, the traditional \cite{Shakura:1973} $\alpha$ parameter; the ratio of Maxwell stress to magnetic energy, $\alpha_{\rm mag}$ \cite[]{Hawley:2011}\footnote{This is closely related to the ``tilt-angle" measured by several authors \citep{Guan:2009a,Beckwith:2011,Sorathia:2011}.}; the ratio of Maxwell to Reynolds stress, ${\rm MR}$; and the ratio of magnetic energy component $i$ to the total magnetic energy, ${\rm BR}_i$. The angled brackets denote a volume average, whereas the overbars indicate a time average. These quantities are shown in Figure \ref{turb_avg} and in Table \ref{tbl:sat}. We find that, within the precision possible given the limited duration of our simulations, all of these metrics appear to converge for domain sizes $L_x \ge 2 H$, i.e. 
for all but the smallest domain. We find that $\alpha$ varies between $0.02-0.03$; $\alpha_{\rm mag} \sim 0.4$, consistent with expectations for converged MRI turbulence \cite[]{Hawley:2011,Sorathia:2011}; the ratio of Maxwell to Reynolds stresses is $\sim 3-4$; while finally ${\rm BR}_x \sim 0.11$, ${\rm BR}_y \sim 0.84$, ${\rm BR}_z \sim 0.05$. These diagnostics can be directly compared with the results of \cite{Beckwith:2011}, who characterized the properties of MRI-driven MHD turbulence in a well-resolved {\em global} simulation. In that work, the authors found that disc-turbulence measured in regions well away from the innermost stable circular orbit was characterized by $\alpha \sim 0.025$ (directly in the middle of the range found here), $\alpha_{\rm mag} \sim 0.3$ and finally ${\rm BR}_x \sim 0.1$, ${\rm BR}_y \sim 0.88$, ${\rm BR}_z \sim 0.02$. We conclude that the basic measures of the turbulence reported here are broadly consistent with those in \cite{Beckwith:2011}.

While boxes larger than $2H\times4H\times8H$ show high levels of convergence in these diagnostics, the behaviour of the smallest box ($\half$) is quite different. While $\alpha = 0.022$ in this case, consistent with the other domain sizes presented here, the other diagnostics are significantly outside of the range of parameter space occupied by the larger domains. Most tellingly, we find that $\alpha_{\rm mag}$ is a factor $2.5$ \emph{smaller} than the expected value for converged MRI turbulence \citep{Hawley:2011,Sorathia:2011}. We have verified this behavior for calculations with the same domain size but with 72 and 144 zones per $H$ (see Table~\ref{tbl:sat}).  The run FT0.5h appears to have values closer to the converged values.  However, analyzing the time history of these quantities show that they are not quite saturated between orbit 50 and the end of the calculation.  In particular, $\alpha_{\rm mag}$ is still decreasing during the period over which we average.  Given limited computational resources, we did not run this case out further, but instead we expect that $\alpha_{\rm mag}$ will saturate at a value lower than that presented here.  The bottom line is the anomalous behavior of the small domain simulations is {\it not} a resolution effect.  

\begin{figure}
\begin{center}
\leavevmode
\includegraphics[width=\columnwidth]{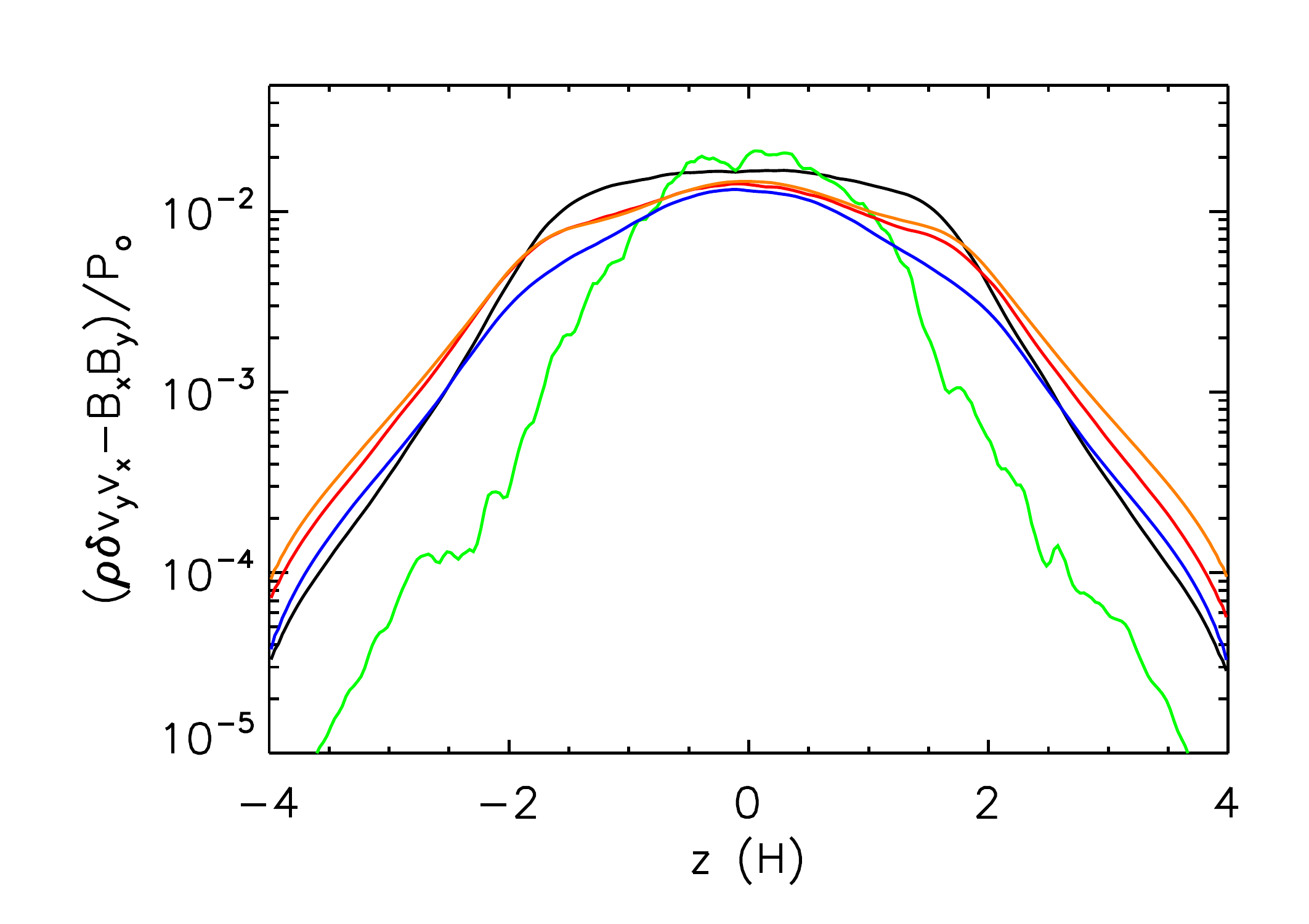}
\includegraphics[width=\columnwidth]{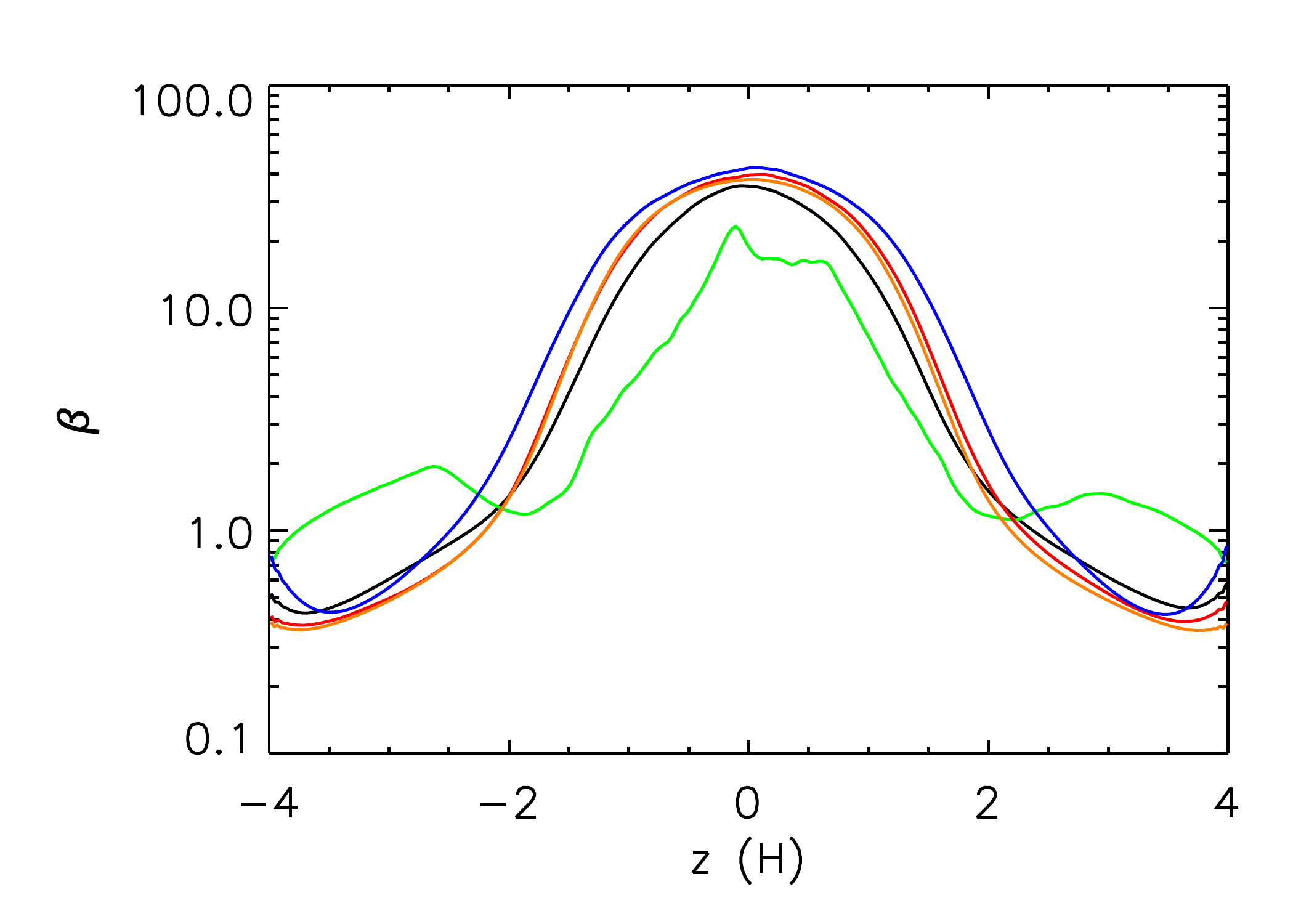}
\end{center}
\caption[]{Time- and horizontally-averaged vertical profiles of the total stress normalized to the initial mid-plane gas pressure (top panel) and gas $\beta = 2P/B^2$ parameter (bottom panel). In both cases, the numerator and denominator are separately averaged.  The green curve is from $\half$, black is $\two$, red is $\four$, blue is $\eight$, and orange is $\sixteeny$. With the exception of the smallest domain, the vertical structure of these quantities are consistent between all calculations: the stress is relatively flat until $|z| \sim 2H$, after which it drops off rapidly, and $|z| \sim 2H$ is also where the $\beta \sim 1$ line is crossed.}
\label{all_z}
\end{figure}


\begin{figure}
\begin{center}
\leavevmode
\includegraphics[width=\columnwidth]{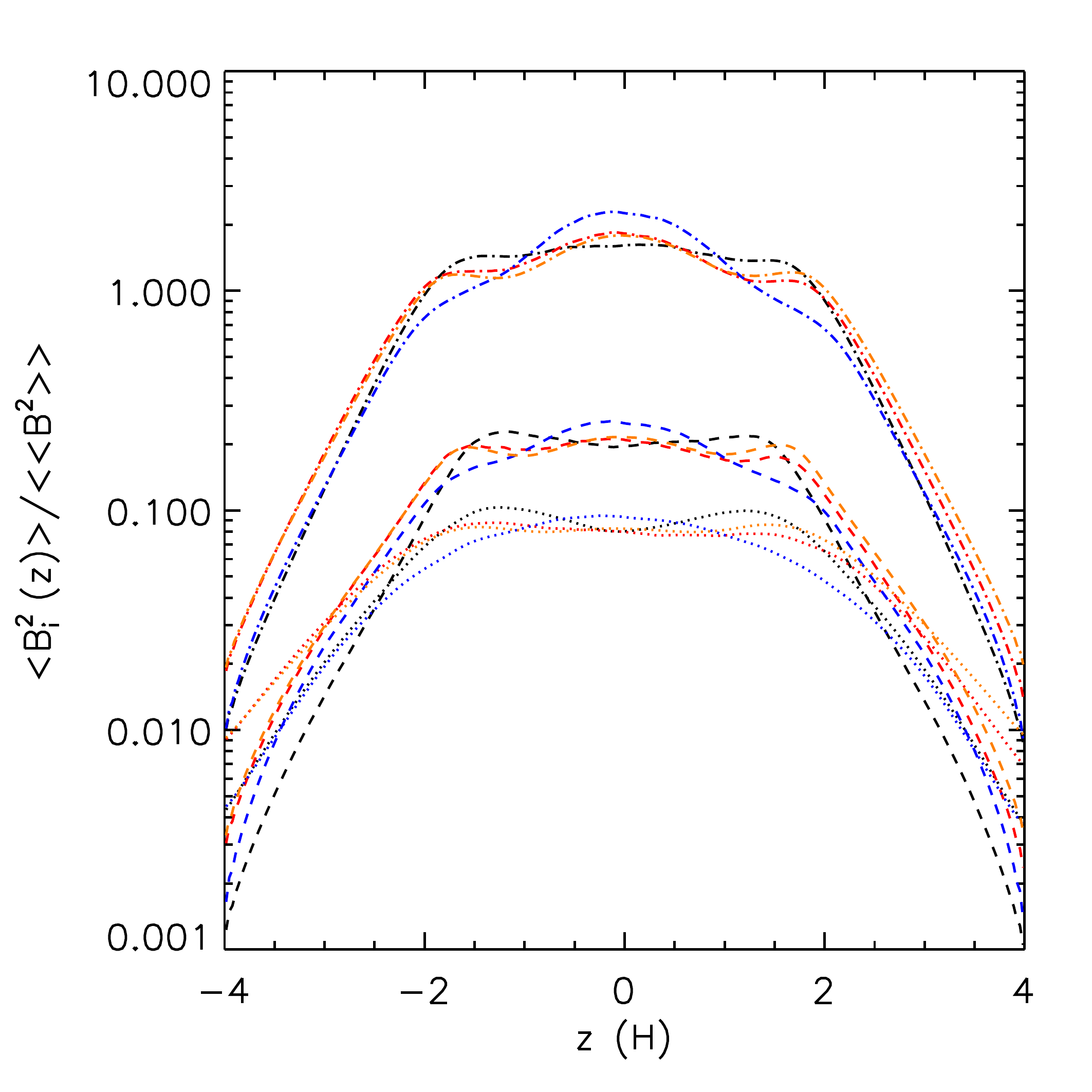}
\end{center}
\caption[]{The time- and horizontally-averaged vertical profile of the magnetic field strength in individual components, $\left< B^2_i \right>$ ($i=x,y,z$), normalized to the volume integrated total magnetic energy strength, $\left<\left< B^2 \right>\right>$. These quantities are separately averaged. We omit data from the smallest domain (\half) for the purposes of clarity. The black curve is from $\two$, red is $\four$, blue is $\eight$, and orange is $\sixteeny$. As with Figure \ref{all_z}, the vertical structure of these quantities are consistent between all calculations: the energy density in toroidal fields dominates throughout the domain. In the region $|z| \lesssim 2H$, the ratio of energy densities is $4:40:1$ ($x:y:z$), while in the region $|z| \gtrsim 3H$, the energy density in vertical fields exceeds that in radial fields, becoming comparable to (but still smaller than) the energy density in toroidal fields.}
\label{topology}
\end{figure}


The turbulence in the quasi-stationary state can be further characterized by the vertical structure of the simulated disc. In \cite{Simon:2011b}, we examined the statistical distribution of turbulent velocity fluctuations as a function of height above the midplane in runs \two, \four, \eight~and \sixteeny, and showed that there was a non-negligible supersonic component for $|z| \gtrsim 3H$ \cite[see Fig. 3 of][]{Simon:2011b}. In run \two ~and \four, this component occupied $\sim1\%$ and $\sim5\%$ of the probability distribution space respectively, while for simulations \eight~and \sixteeny, the contribution of supersonic velocity fluctuations converged at $\sim10\%$ of the probability distribution space, highlighting the possibility of \emph{observable} differences between local and mesoscale treatments of  magnetized accretion disc turbulence.

Fig. \ref{all_z} shows the vertical structure of the total stress normalized to the initial mid-plane gas pressure and the gas $\beta = 2P/B^2$ parameter in the quasi-stationary state. Here, the numerator and denominator for these quantities are time- (from orbit 50 onward) and horizontally-averaged separately before computing their ratio. With the exception of the smallest domain run, the vertical structure appears reasonably consistent between the various domain sizes; the stress is relatively flat until $|z| \sim 2H$, after which it drops off rapidly, and $|z| \sim 2H$ is also where the $\beta \sim 1$ line is crossed. The smallest domain again appears pathological in comparison to the larger domain sizes: the stress is sharply concentrated towards the midplane, where the disc is weakly magnetized ($\beta \sim 20$ cf. $\beta \sim 40$ in the larger domains), while $\beta \lesssim 1$ for $|z| \gtrsim 2H$.


Fig.~\ref{topology} shows the vertical profile of the energy density in toroidal, radial and vertical magnetic fields normalized to the total magnetic energy density, e.g. $\left< B^2_i \right>/\left<\left< B^2 \right>\right>$ ($i=x,y,z$). As before, we average these quantities separately before taking their ratio and we omit data from the smallest domain (run \half) for the purposes of clarity. As was the case with the data shown in Figure \ref{all_z}, the vertical structure of these quantities appear to be reasonably consistent between the various domain sizes: the energy density in toroidal fields dominates throughout the domain. In the region $|z| \lesssim 2H$, the ratio of energy densities is $4:40:1$ ($x:y:z$), while in the region $|z| \gtrsim 3H$, the energy density in vertical fields exceeds that in radial fields, becoming comparable to (but still smaller than) the energy density in toroidal fields. This leads us to an interesting point: the largest domain (\sixteeny) has vertical magnetic field strengths at $|z| = 4H$ that exceed the strength of the radial magnetic field here by a factor $\sim2$ and are within a factor of $\sim2$ of the toroidal magnetic field strength. Given the vertical gradients of each component of the magnetic field strength, it appears that the energy density in vertical fields will exceed that in toroidal magnetic fields by $|z| = 6H$; such a configuration could lead to angular momentum transport loss through a magnetically launched wind \cite[e.g.][]{Blandford:1982}. Repeating run $\sixteeny$ with an extended \emph{vertical} domain (i.e. a domain size $16H\times32H\times16H$) could therefore yield insight into the launching of a magnetically driven wind from a turbulent accretion disc. The computational cost of such a simulation means that it is beyond the scope of the present work.
\vspace{-0.2in}


\begin{figure*}
\begin{center}
\leavevmode
\includegraphics[width=\textwidth]{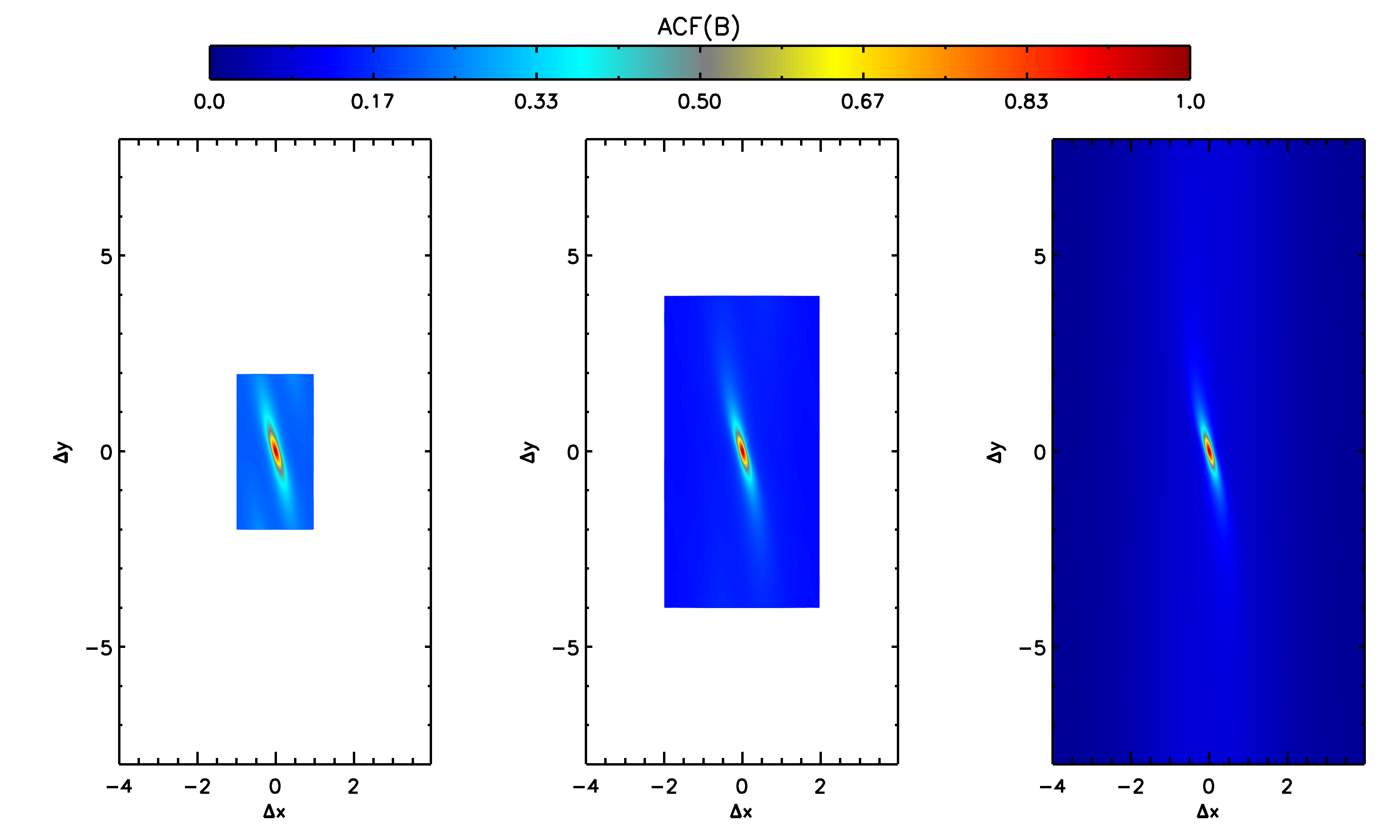}
\end{center}
\caption{ The ACF of the magnetic field, as defined by equation~(\ref{corr_func}) and calculated for $|z| \le 2H$, for $\two$ (left), $\four$ (middle), and
$\eight$ (right) in the $\Delta z = 0$ plane.  The tilted centroid structure of the magnetic field is the same size and shape in all three domains.}
\label{corr2d_b_long_zle2H}
\end{figure*}

\begin{figure}
\begin{center}
\leavevmode
\includegraphics[width=\columnwidth]{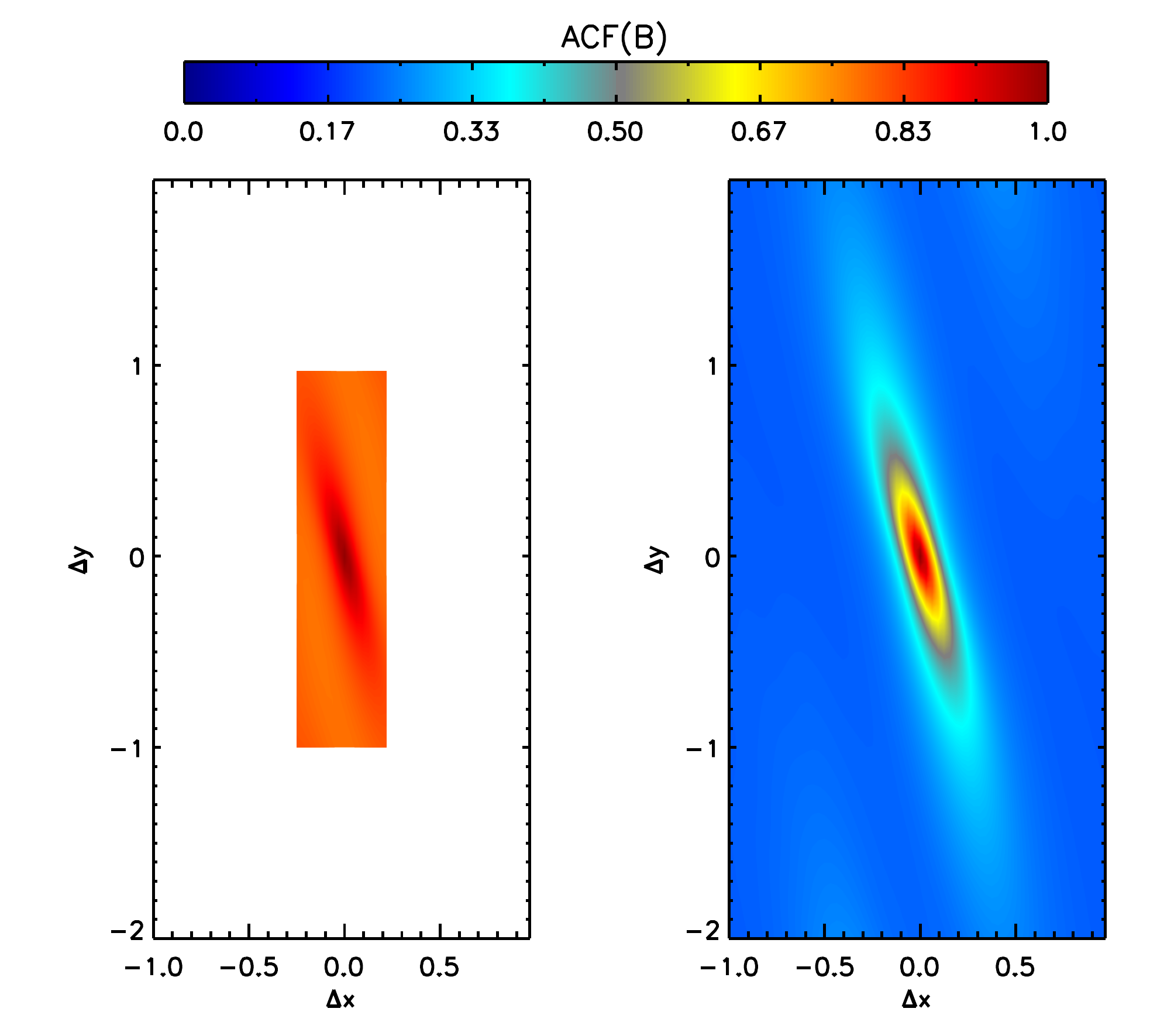}
\end{center}
\caption{The ACF of the magnetic field, as defined by equation~(\ref{corr_func}) and calculated for $|z| \le 2H$, for the smallest domain ($\half$, left) and
the next largest domain ($\two$, right) in the $\Delta z = 0$ plane. While the tilted centroid structure of the magnetic field appears to be mostly contained within the larger box, it is most
definitely not contained in the smaller domain.}
\label{corr2d_b_long_smaller}
\end{figure}

\section{Mesoscale Spatial Correlations}\label{mesoscale}

\subsection{Magnetic Field}\label{field}

To examine the structure of the turbulence in the ($x,y$) plane, we employ the autocorrelation function (ACF) \cite[see][]{Guan:2009a}, defined as,
\begin{equation}
\label{corr_func}
{\rm ACF}(f(\bm{\Delta x})) \equiv \left\langle \frac{\int f(t,\bm x) f(t,\bm x + \bm{\Delta x}) d^3{\bm x}}{\int f(t,\bm x)^2 d^3{\bm x}}\right\rangle,
\end{equation}
\noindent where $f$ is the fluid quantity of interest, and the brackets denote a time-average. Note that we have defined the ACF to be normalized by its maximum value (at $\Delta x = \Delta y = \Delta z = 0$), and unlike \cite{Guan:2009a},
we do not subtract off any mean quantities before calculating the ACF.  Furthermore, for vector quantities, such as the magnetic field, we take the ACF of each component and then sum them as was done in \cite{Guan:2009a}; ACF($B$) = ACF($B_x$) + ACF($B_y$) + ACF($B_z$).  The time average is done from orbit 50 to 125 in all cases.

Because of the change in vertical structure at $|z| \sim 2H$,  we have created two sets of ACFs, one for which we restricted the calculation to $|z| \le 2H$, the other of which is for $|z| > 2H$. This will allow us to probe what has been referred to as the turbulent ``disc region" ($|z| \lesssim 2H$) separately from the ``coronal region" ($|z| \gtrsim 2H$). First focusing on the disc region,  Fig.~\ref{corr2d_b_long_zle2H} shows the ACF of the magnetic field at $\Delta z = 0$ for the runs $\two$, $\four$, and $\eight$.  
To account for the large change in scale, we plot the same quantity for the two smallest domains, \half and \two, in Fig.~\ref{corr2d_b_long_smaller}. 


As in the vertical structure plots, there is a striking difference between the smallest domain and the larger ones.  Although for the larger domains, the ACF has a small but nonzero value over extended regions (a point that we shall return to shortly), it is obvious that the
tilted and elongated {\em centroid} of the ACF is well contained within the domain.  This is not the case for \half; indeed, it appears that the correlation length of ${\bm B}$ extends beyond the size of the region, and that the peak of the centroid is not too much larger than the value of the ACF away from the centroid. Since $\alpha_{\rm mag}$ is proportional to the "tilt angle" away from the $y$ axis \citep{Guan:2009a}, the low average value of $\alpha_{\rm mag}$ along with the extended ACF imply that transport of angular momentum by magnetic fields is reliant on the action of strong coherent toroidal magnetic fields, rather than small-scale turbulent structures. 

\begin{figure}
\begin{center}
\leavevmode
\includegraphics[width=\columnwidth]{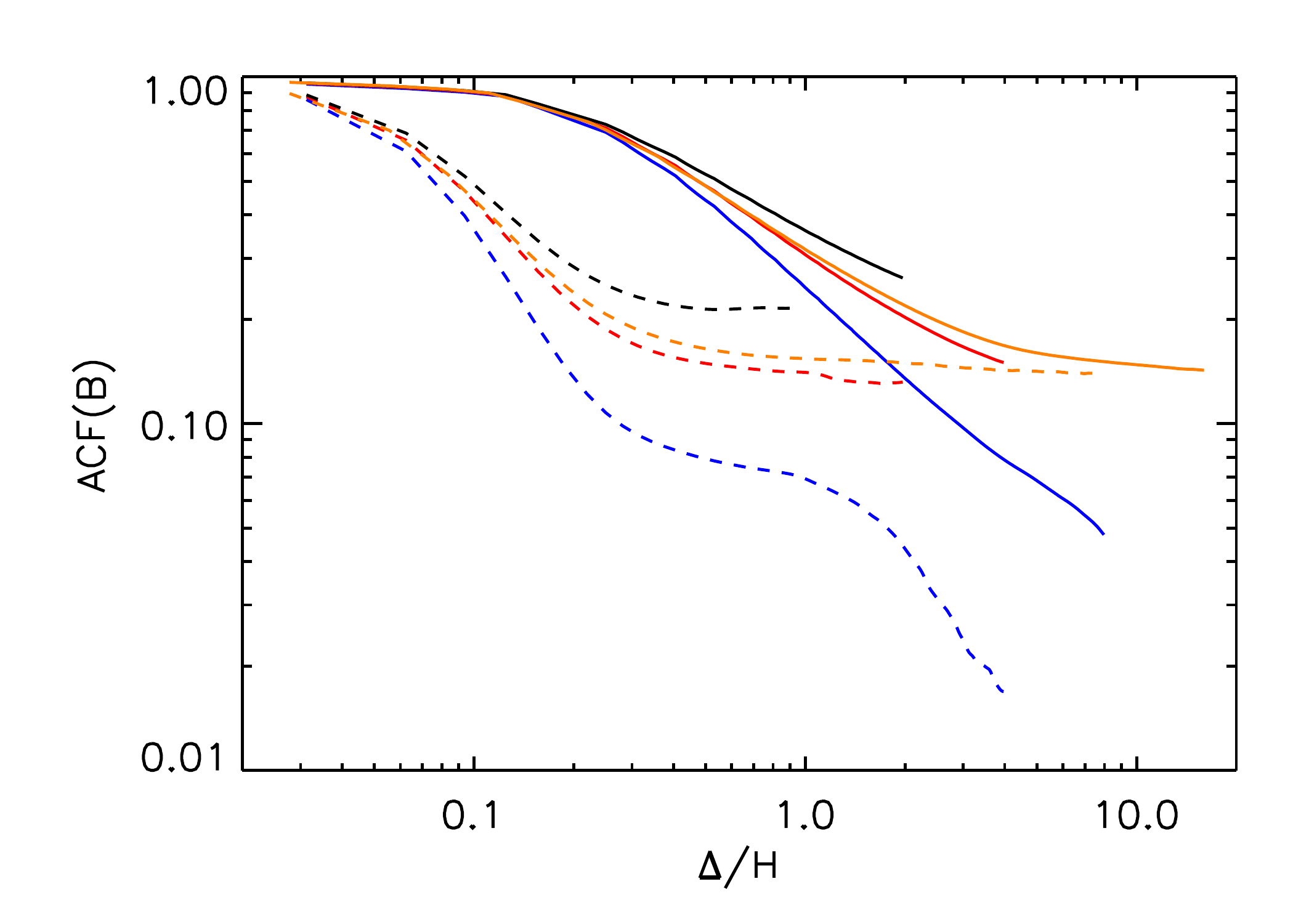}
\includegraphics[width=\columnwidth]{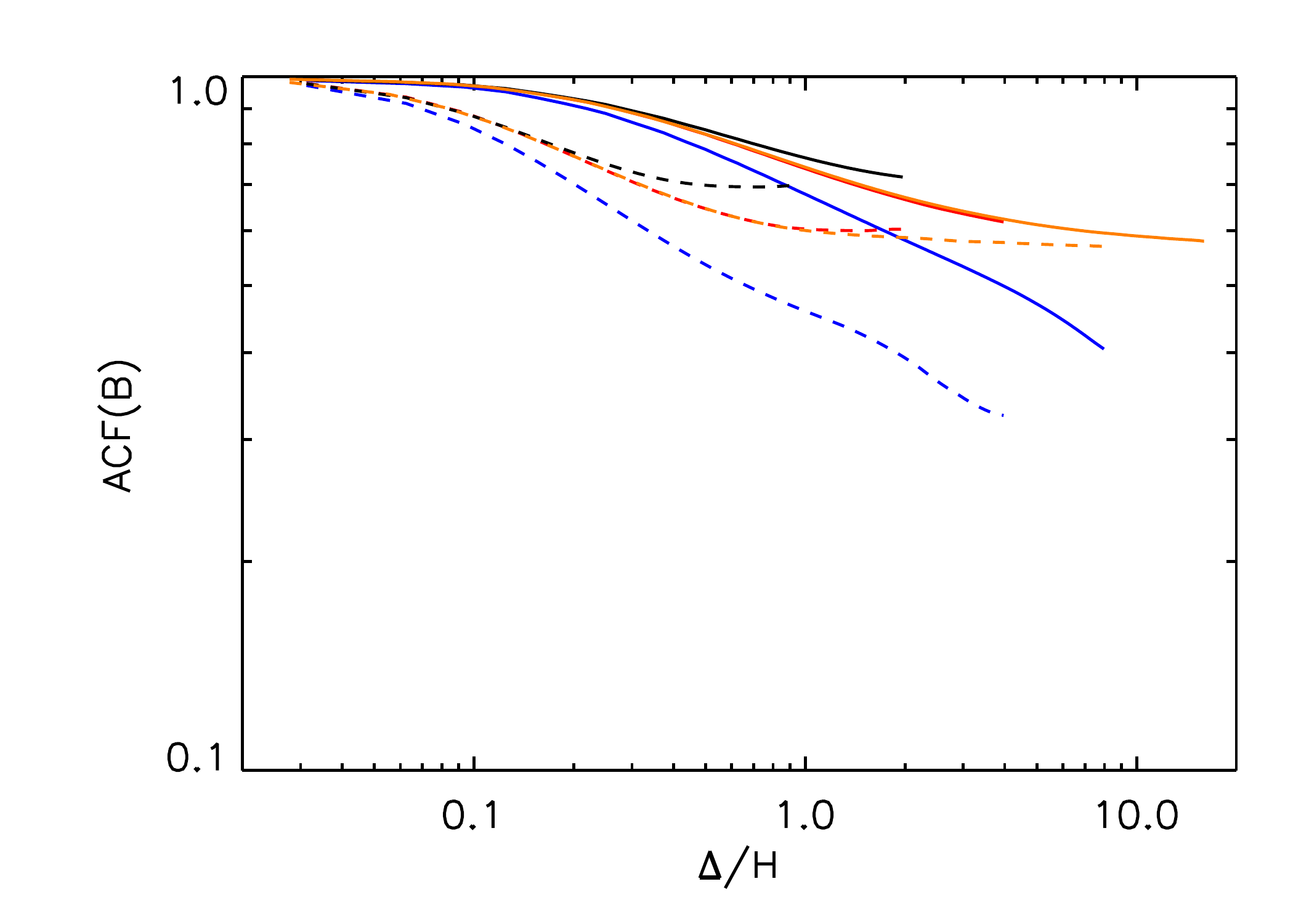}
\end{center}
\caption{The ACF defined by equation~(\ref{corr_func}) for the magnetic field taken along the major (solid lines) and minor (dashed lines) axes of the centroid.  The top plot is ACF($B$) calculated for $|z| \le 2H$, whereas the bottom plot is the same quantity but for $|z| > 2H$. We omit data from the smallest domain (\half) for the purposes of clarity. 
 The black curve is from $\two$, red is $\four$, blue is $\eight$, and orange is $\sixteeny$. The magnetic field in the coronal regions has a larger correlation length than that within $2H$ of the mid-plane.  In all cases, the magnetic field have a strongly localized component as well as a non-zero extended component that fills the simulation volume.}
\label{corr1d}
\end{figure}

\begin{figure}
\begin{center}
\leavevmode
\includegraphics[width=\columnwidth]{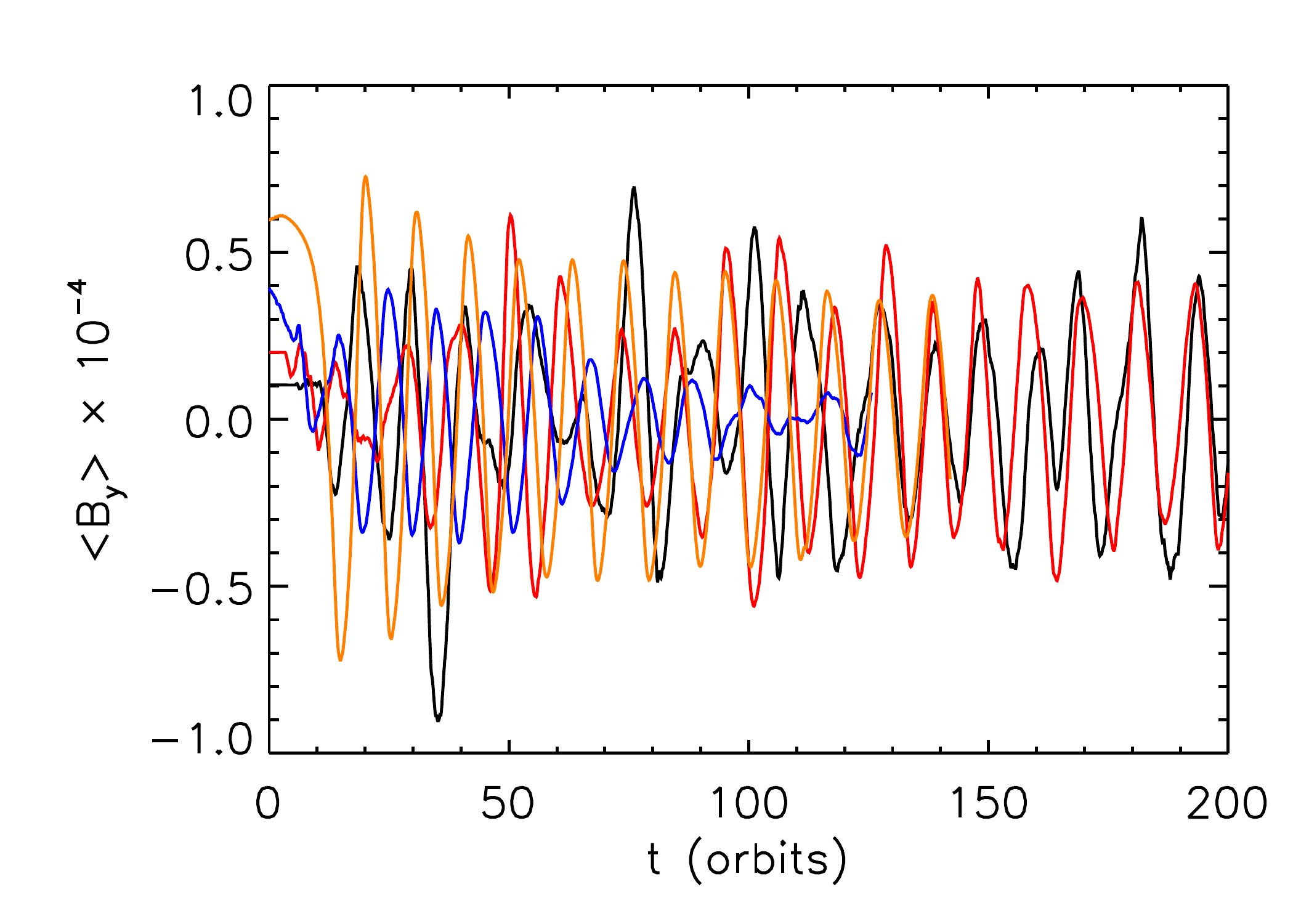}
\end{center}
\caption{Time evolution of the volume averaged toroidal field (for all $x,y$ and for $|z| \le 2H$) in code units. $\two$ is represented by the black curve, $\four$ is the red curve, $\eight$ is the blue curve, and $\sixteeny$ is the orange curve.  All simulations show the oscillation of the mean toroidal field, but the oscillation amplitude is lower in $\eight$.}
\label{by_t}
\end{figure}

Beyond the smallest domain, there generally  appears to be two components to the ACF, one of which is strongly localized and tilted and which does not change significantly with domain size.  The other component is more uniformly distributed in $(\Delta x, \Delta y)$ space and represents the background magnetic field.   This component fills the entire domain, and
in that sense cannot converge with increasing domain size. That is, as one makes the shearing box larger and larger, there is magnetic field that is always volume-filling, and thus there is always a component to the field structure that is correlated on the largest scales of the box.

To demonstrate this more quantitatively,  Fig.~\ref{corr1d} shows 1D slices of the ACF along the major and minor axes of the tilted structure.  These plots include the largest box ($\sixteeny$), but 
exclude the smallest domain size. The lower plot corresponds to ACF($B$) calculated in the coronal region The ACF calculation in this region was done
for $z > 2H$ and $z < -2H$ separately, and the resulting ACFs were averaged together.  These plots again demonstrate that there are two components to the ACF, one that is concentrated and tilted, the other of which is uniform and extended.  The tilted correlation structures are significantly larger in the coronal region compared to the disc region; the magnetic field has a larger correlation length in the coronal region.  Furthermore, the value of the ACF for the extended component is larger than the corresponding component in the disc region.  These observations suggest that the mean background field becomes a greater fraction of the magnetic energy in the coronal region \cite[this is consistent with the two-dimensional correlation functions of][]{Guan:2011} and that the MRI modes that do operate in these regions produce fluctuations of relatively large length scale.  This latter point is likely a result of the characteristic MRI wavelength becoming larger as the density drops (i.e., the $\alf$ speed increases) in the corona.


\begin{figure*}
\begin{center}
\leavevmode
\includegraphics[width=\textwidth]{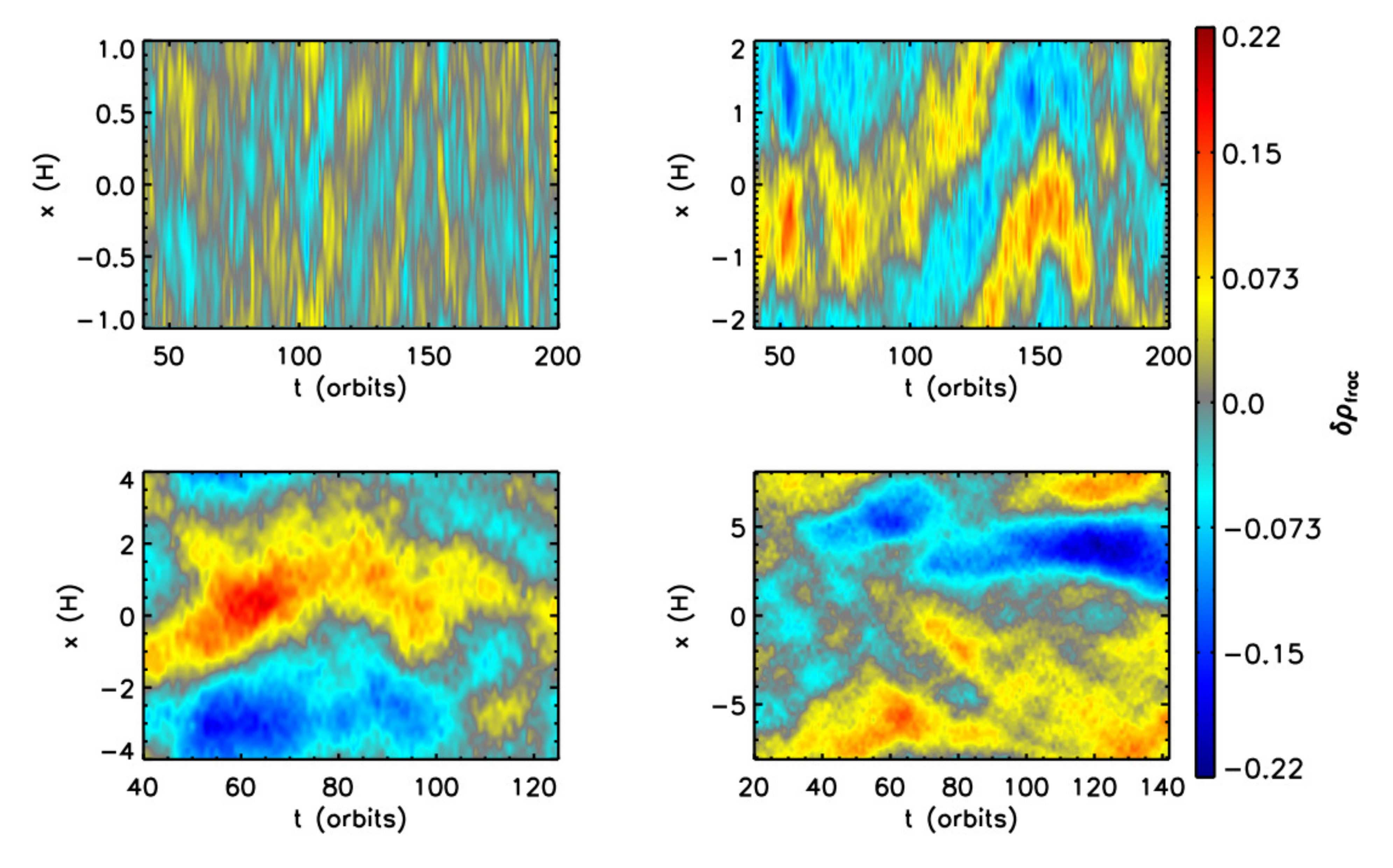}
\end{center}
\caption{Space-time diagram of fractional gas density fluctuations, $\delta\rho_{\rm frac}$ (as defined in the text) in the $(t,x)$ plane.  Note the different scales on both axes in each plot.  For $\four$ (upper right),
and $\eight$ (lower left), there exists a $k_x L_x/(2\pi) = 1$ mode zonal flow.  The zonal flow in $\sixteeny$ (lower right) has a structure more complex
than a simple $k_x L_x/(2\pi) = 1$ mode.}
\label{sttx_d}
\end{figure*}


\begin{figure*}
\begin{center}
\leavevmode
\includegraphics[width=\textwidth]{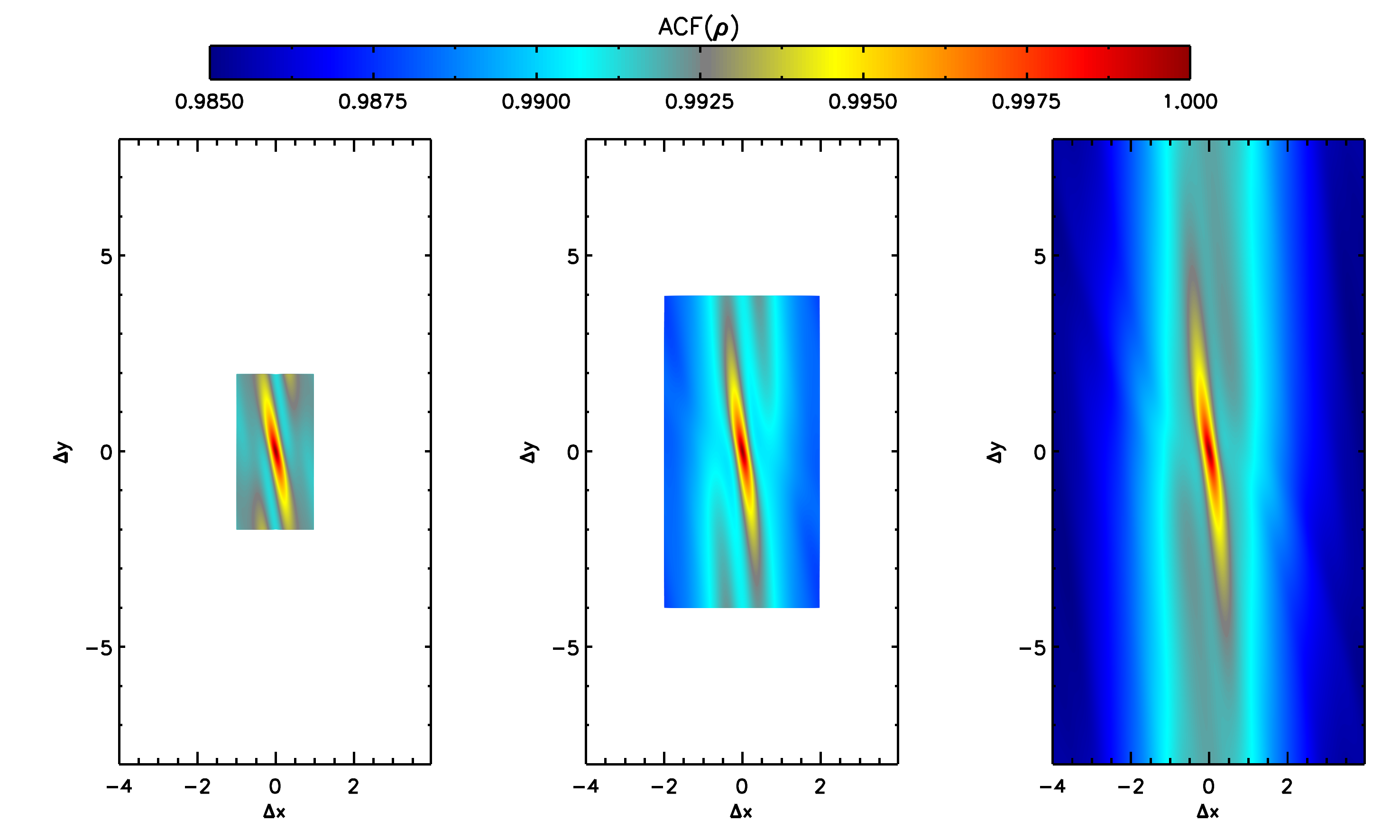}
\end{center}
\caption{The ACF of the gas density as defined by equation~(\ref{corr_func}) for $\two$ (left), $\four$ (middle), and
$\eight$ (right) in the $\Delta z = 0$ plane. In all three ACFs, there is a tilted centroid structure.  However, as domain size is increased, a larger scale
axisymmetric structure appears.}
\label{corr2d_d_long_zle2H}
\end{figure*}

Beyond comparing the coronal vs. disc regions, it is worth noting the convergence properties of the ACF as the domain size is increased.  In both the coronal and disc regions,
the inner centroid appears to be convergent with domain size.  While the different ACFs at large $\Delta/H$ do not lie on top of each other, they are reasonably consistent with each other, with the exception of $\eight$ (the blue curve).   In both the corona and disc regions, the extended ACF of $\eight$ appears significantly lower than the other domain sizes.   We have examined the evolution of a volume-averaged $\langle B_y \rangle$ for these calculations (shown in Fig.~\ref{by_t}); the $\eight$ run shows smaller amplitude (though still regular) variations in $\langle B_y \rangle$, resulting in a smaller background field component relative to small scale turbulent structures. The smaller background field relative to the turbulent fluctuations would certainly account for the lower extended ACF for $\eight$.  We are not entirely sure why the background field is weaker in $\eight$ in the first place, though some statistical variation is to be expected since the evolution of $\langle B_y \rangle$ is modulated on rather long timescales \citep{Simon:2011a}.

To summarize, the horizontal structure of the magnetic field consists of an inner centroid that is tilted with respect to the azimuthal direction, as well as an extended component, likely due to the background magnetic field.  The general shape of the ACFs converge, though as domain size is increased, there will always be structure at the largest scales of the box. These results have consequences for the physics of angular momentum transport in magnetized accretion discs, which we discuss in detail in \S\ref{stress}.

\subsection{Zonal Flows}\label{density}

We now turn to the structure and evolution of the gas density in the simulated discs. We are particularly interested in the existence and properties of zonal flows, long lived axisymmetric perturbations to the gas density that result from a geostrophic balance between pressure and Coriolis forces. \cite{Johansen:2009} identified such flows in simulations of MRI disc turbulence, and suggested that 
they occur as a result of an inverse cascade in magnetic energy, which induces regions of super- and sub-Keplerian velocities. In broad terms, zonal flows are of physical interest because their existence within protoplanetary discs would provide a way to trap solid particles that are partially coupled to the gas disc through aerodynamic forces. 
Here, following prior work by \cite{Johansen:2009} and \cite{Yang:2011}, we study whether the size and scale of the zonal flows converges with domain size.

Fig.~\ref{sttx_d} shows space-time diagrams in the $(t,x)$ plane of the fractional perturbation $\delta\rho_{\rm frac}$ to the density. 
In calculating $\delta\rho_{\rm frac}$, we average the density over $y$ and $z$, and then subtract off and normalize by the volume averaged $\rho$ at each time.   Note the change in both horizontal and vertical scale in each panel of the figure.  For $\two$ (upper left), there is essentially no indication of a zonal flow, but for the larger boxes, there are large scale long-lived features in $\rho$.  For $\four$ and $\eight$, these features always fill the largest radial scale in the box (i.e., a $k_x L_x/(2\pi) = 1$ mode).  The same diagram for the largest box (lower right), however, shows a structure more complex
than a simple $k_x L_x/(2\pi) = 1$ mode.

\begin{figure}
\begin{center}
\leavevmode
\includegraphics[width=\columnwidth]{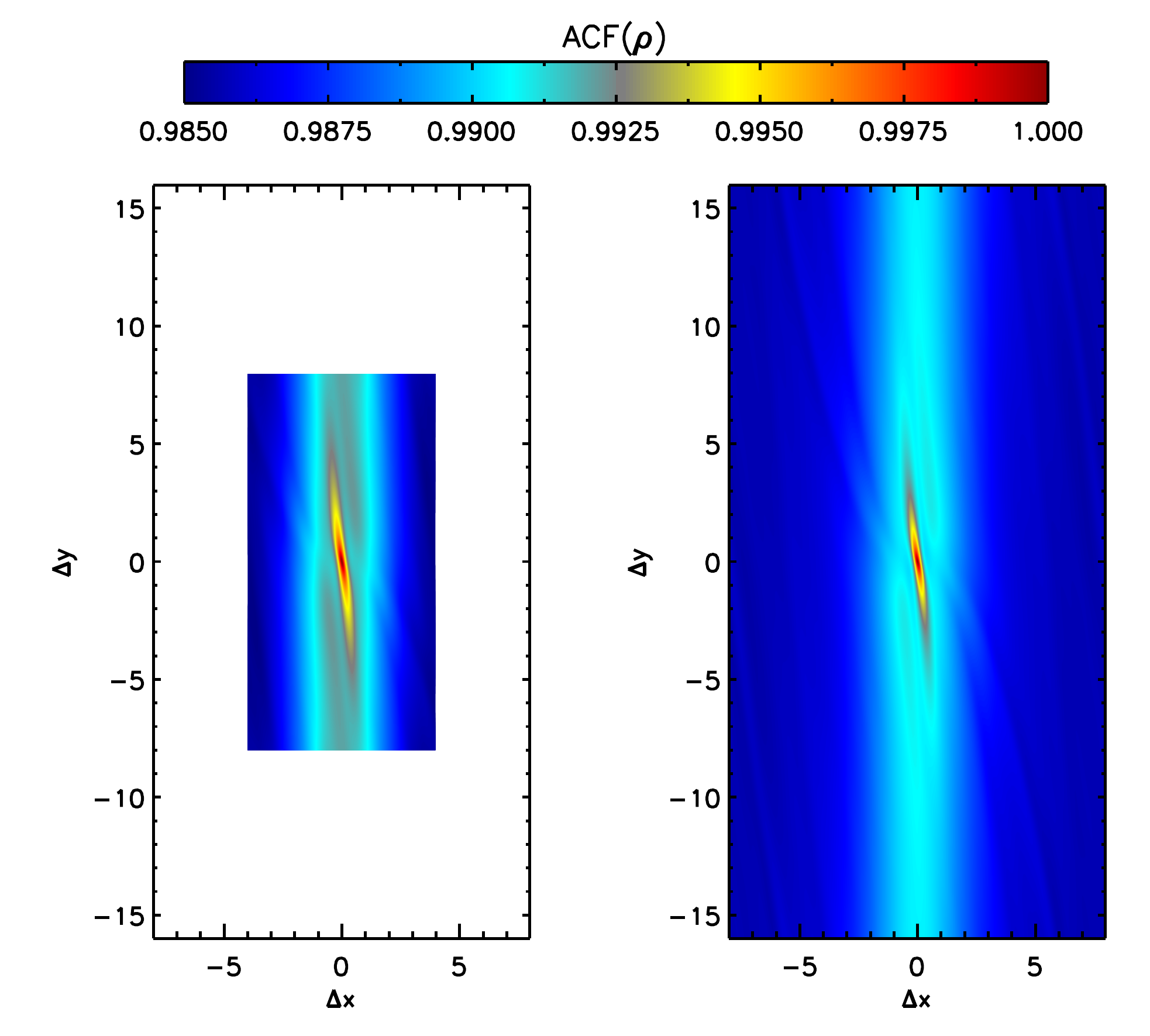}
\end{center}
\caption{The ACF of the gas density as defined by equation~(\ref{corr_func}) for $\eight$ (left) and 
$\sixteeny$ (right) in the $\Delta z = 0$ plane. The density structure is consistent between these two domain sizes, including the large
scale axisymmetric component.}
\label{corr2d_d_long_larger}
\end{figure}

\begin{figure}
\begin{center}
\leavevmode
\includegraphics[width=\columnwidth]{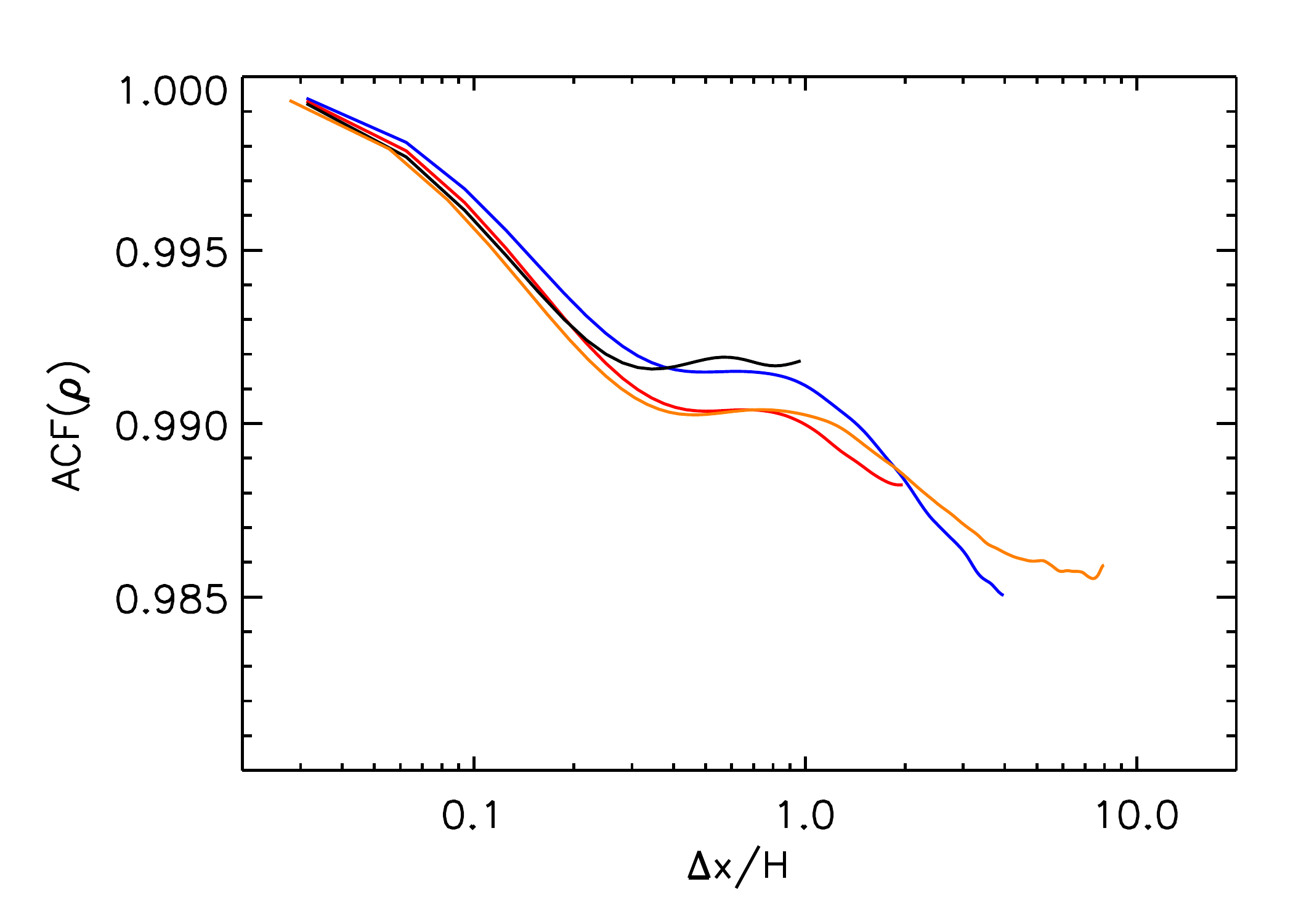}
\end{center}
\caption[]{The ACF of the gas density as defined by equation~(\ref{corr_func}) versus $\Delta x$ along $\Delta y = 0$ and $\Delta z = 0$. The colors denote the various runs; black is $\two$, red is $\four$, blue is $\eight$, and orange is $\sixteeny$. All ACFs show a central component that flattens out at $\sim 0.3 H$.  For boxes larger than $\two$, there is a second dip corresponding to the presence of a zonal flow.  Finally, there is hint of this second dip flattening out for $\sixteeny$, thus providing tentative evidence for an outer scale to the zonal flow that is contained within the simulation domain.}
\label{corr1d_xy_d_long_zle2H}
\end{figure}

To quantify the density structure further, we examine the ACF of $\rho$ in the $(\Delta x, \Delta y)$ plane (the same averaging is done here as for ACF($B$)).  For simplicity, we only calculate the ACF over $|z| \le 2H$. The result for $\two$, $\four$, and $\eight$ is shown in Fig.~\ref{corr2d_d_long_zle2H}, and the ACFs for $\eight$ and $\sixteeny$ are shown in Fig.~\ref{corr2d_d_long_larger}.  Note the range in ACF values from the color bar; in general, the density is very nearly uniform in the horizontal plane.  However, in all domain sizes, there is a tilted and concentrated structure resembling the magnetic field ACF.  This particular structure was also seen in the calculations of \cite{Guan:2009a}. As domain size is increased, an axisymmetric structure appears.  This structure, which appears to be more or less converged between the two largest runs, is due to the presence of zonal flows.  We have also examined a one-dimensional slice along $\Delta y = 0$, shown in Fig.~\ref{corr1d_xy_d_long_zle2H}. The evidence for convergence is less persuasive in this one-dimensional slice than in Fig.~\ref{corr2d_d_long_larger}, and the most that we can conclude is that there are tentative signs that the outer radial scale of the zonal flow is $\Delta x \sim 6H$ (implying a zonal flow wavelength of $\sim 12H$) and is contained within the largest spatial domain modeled.  Running an even larger domain simulation could potentially nail down this radial scale.  However, such large domains become prohibitively computationally expensive given our current resources.

Finally, to explore whether the existence of zonal flows is sensitive to the 
 initial conditions, we ran two additional simulations.  They are the equivalent of $\four$ but initialized with two flux tubes in one case,
and with a uniform toroidal field (at constant $\beta$) in the second case. In both cases, a $k_x L_x/(2\pi) = 1$ mode zonal flow appears around 10-20 orbits into the simulations.

\subsection{Locality of Angular Momentum Transport}\label{stress}

\begin{figure}
\begin{center}
\leavevmode
\includegraphics[width=\columnwidth]{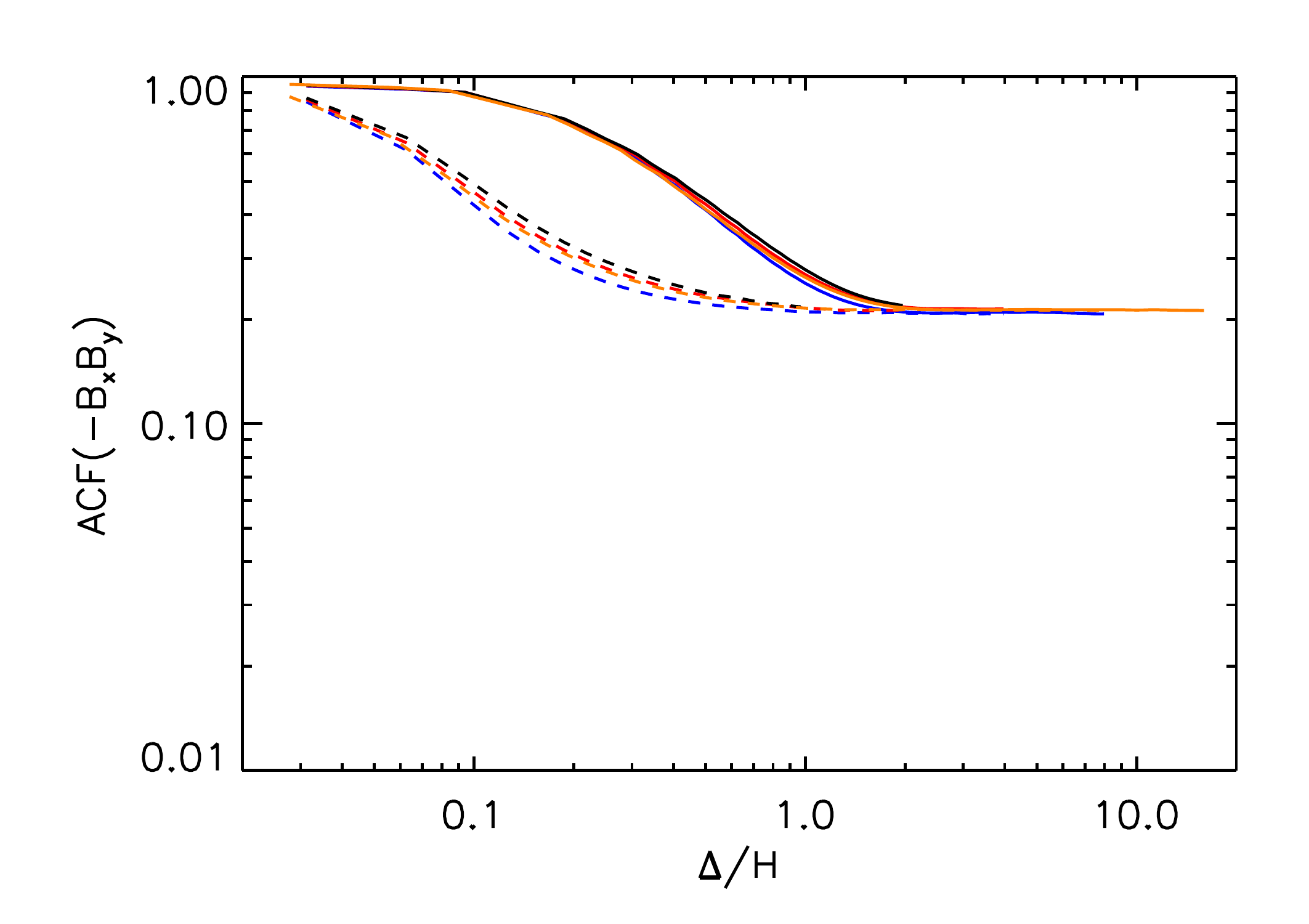}
\includegraphics[width=\columnwidth]{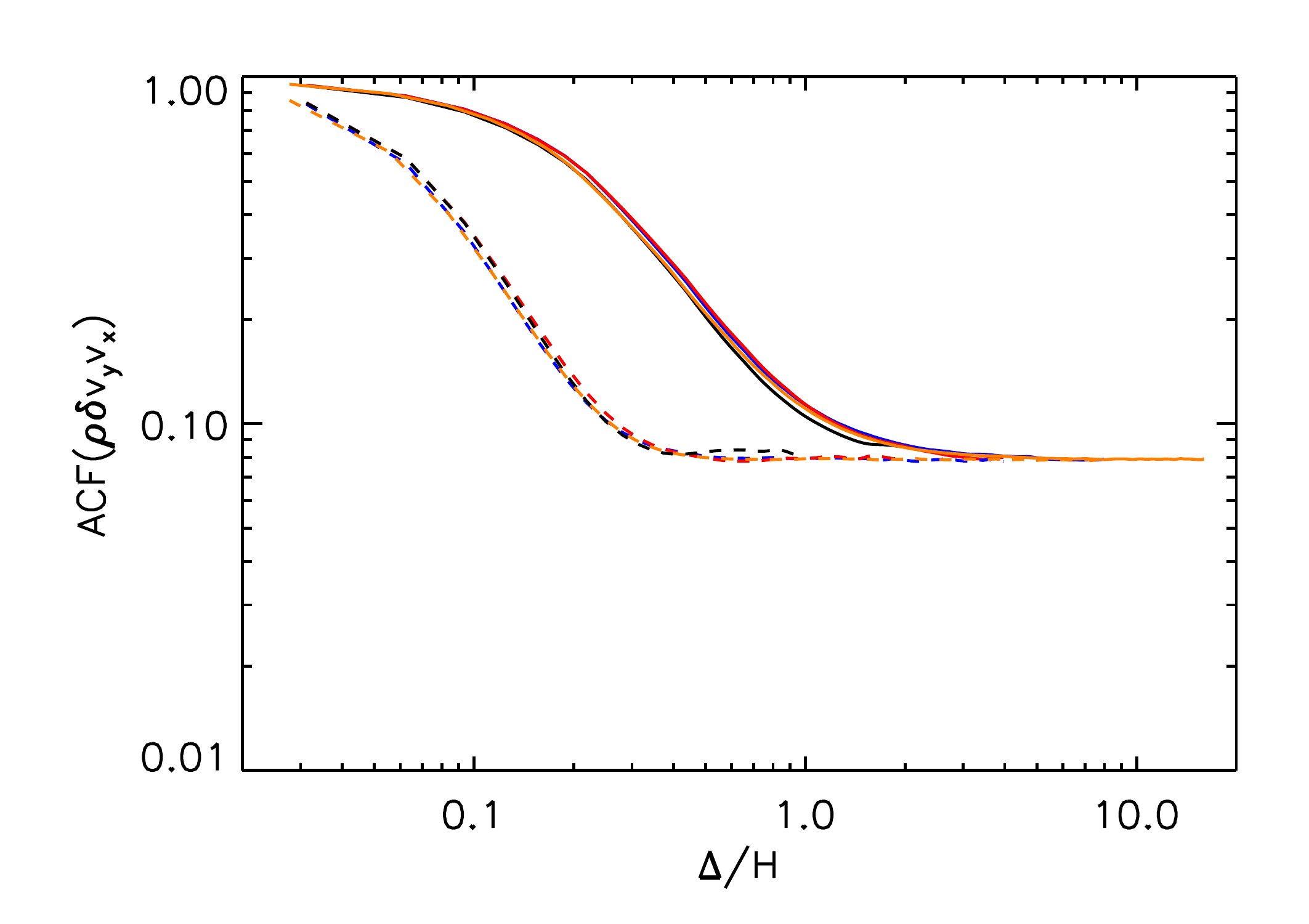}
\end{center}
\caption{As in the top panel of Figure \ref{corr1d} for the ACF of the Maxwell (top panel) and Reynolds (bottom panel) stress. While the ACFs are strongly concentrated within $\lesssim H$ of the centroid, there is an extended component, correlated at $\sim20\%$ for the Maxwell stress and $\sim8\%$ for the Reynolds stress, that fills the simulation domain in all cases.}
\label{stress1d}
\end{figure}

In the preceding sections, we have demonstrated that there are mesoscale correlations within the turbulent disc ($|z|<2H$) in both the magnetic field and the density. The existence of such structures can have important consequences for the physics of angular momentum transport in magnetized accretion discs. For example, if the mesoscale correlations in the magnetic field within the turbulent disc discussed in \S\ref{field} are actively generated from the magnetorotational instability, then these fields can also transport angular momentum through the Maxwell stress term in the angular momentum conservation equation \citep{Balbus:1999}. Similarly, the zonal flows discussed in \S\ref{density} are associated with sub- and super-Keplerian velocities which form due to the action of magnetic forces \citep{Johansen:2009}; such flows can then be associated with transport of angular momentum as they drift through the disc. In both of these cases, we would expect to measure angular momentum transport (through the Maxwell and Reynolds stress, respectively) on scales $\gg H$. Such a result would undercut the assumption that angular momentum transport in magnetized discs can be treated as a purely local process \citep{Shakura:1973}.

We can test these ideas by examining the ACF for the Maxwell and Reynolds stress. If angular momentum transport within the disc is determined locally, we would expect the ACF of the stress to decay rapidly on scales $\gtrsim H$; correlations on scales $\gg H$ would indicate a \emph{non-local} component to angular momentum transport \citep{Gammie:1998}. Figure \ref{stress1d} plots one-dimensional slices along the major and minor axes of the tilted centroid component of the ACF for the Maxwell stress, ${\rm ACF}(-B_x B_y)$ (top panel), and Reynolds stress, ${\rm ACF}(\rho v_x \delta v_y)$ (bottom panel), calculated over the region $|z| \le 2H$, i.e. within the ``turbulent disc''. The data of this figure demonstrate that while the ACFs are strongly concentrated within $\lesssim H$, there is an extended component, correlated at $\sim20\%$ for the Maxwell stress and $\sim8\%$ for the Reynolds stress, that fills the simulation domain. Essentially, these results imply that there is a net background Maxwell and Reynolds stress resulting from mesoscale structures in the magnetic field and density, on top of which turbulent stress fluctuations are imposed.

To verify this result, we (for all runs except $\half$) examined the time-averaged (from orbit 50 onward) vertical profile of the quantity 

\begin{equation}
\frac{-\langle B_x\rangle \langle B_y\rangle}{\langle-B_xB_y\rangle}, 
\end{equation}

\noindent
which represents the fraction of the average Maxwell stress that resides in the large scale background magnetic field.  We find that this ratio is quite small near the disk mid-plane but increases dramatically away from the mid-plane, leveling out to $\sim 0.3-0.4$ for $|z| \gtrsim 2H$.  The conclusion from these calculations is clear: while angular momentum transport within magnetized accretion discs is highly localized, there is, in addition, a significant \emph{non-local} component, the existence of which implies that models founded upon a purely local description of angular momentum transport are incomplete.

The origin of these mesoscale correlations in the Maxwell stress can be understood, at least in part, by considering the saturation characteristics given in Table \ref{tbl:sat}. In all of the simulations, the turbulent disc is dominated by strong toroidal magnetic fields (making up $\sim80\%$ of the magnetic energy in simulation $\sixteeny$). The toroidal field MRI is well-resolved according to the criteria of \cite{Sano:2004}, which specifies the number of zones per characteristic wavelength of the instability.  In principle, all modes with wavelengths \emph{longer} than this characteristic wavelength are unstable. This is in contrast to the case of the vertical field MRI, where wavelengths longer than that of the pressure scale height are expected to be stabilized.  Thus, it is not unexpected that long wavelength structures form in the turbulence from the toroidal field MRI.  We discuss this further in Section~\ref{discussion}.


\section{Temporal Variability}\label{var}

\begin{figure*}
\begin{center}
\leavevmode
\includegraphics[width=\textwidth, viewport=220 175 1600 1020,clip]{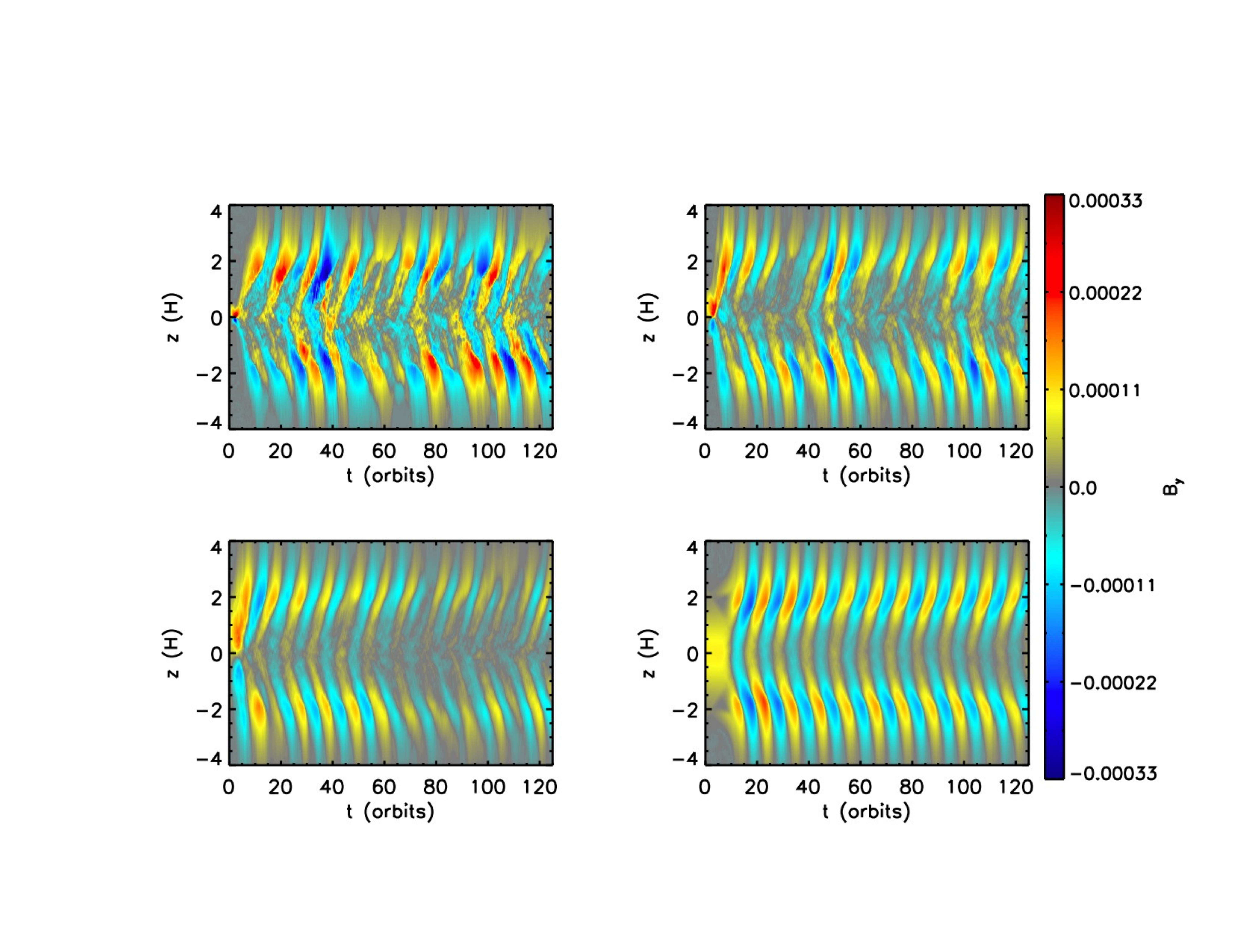}
\end{center}
\caption[]{Space-time diagram in $(t,z$) for the horizontally averaged $B_y$ in $\two$ (upper left), $\four$ (upper right), $\eight$ (lower left), and $\sixteeny$ (lower right).  The so-called "butterfly diagram" present
in vertically stratified MRI-driven discs is apparent in all four domain sizes, and becomes more regular as the domain size is increased.
}
\label{sttz_by}
\end{figure*}

We now return to the behaviour of the temporal variability in domains of different size. From 
Fig.~\ref{stress2H}, it is clear that there is a marked decrease in temporal variability as the domain size is increased.  
This is generically true for a variety of quantities. 
Figure~\ref{sttz_by}, for example, shows the ($t,z$) space-time diagram of $B_y$ for the four largest domain sizes.  
The dominant feature in this diagram is the $\sim$10 orbital period
flipping of $B_y$ that has been observed in many previous calculations \cite[e.g.,][]{Brandenburg:1995,Davis:2010,Simon:2011a,Guan:2011}. However, superimposed upon this is a stochastic component to $B_y$, which becomes smaller as the domain size is increased.

\begin{figure}
\begin{center}
\leavevmode
\includegraphics[width=0.95\columnwidth]{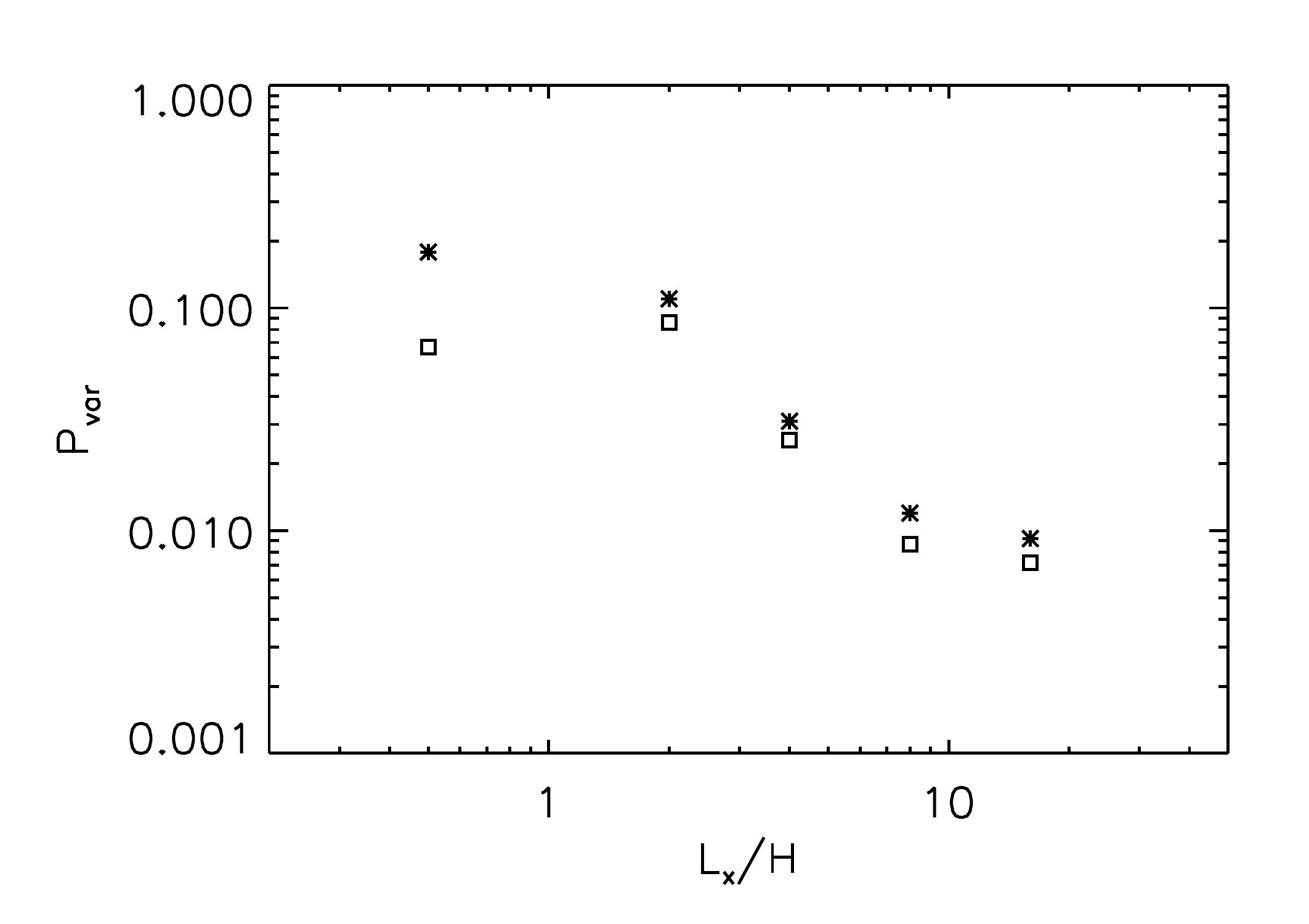}
\end{center}
\caption[]{Variability of the volume-averaged stress (squares) and magnetic energy (asterisks), as defined by $\pv$ (equation~\ref{pvar}) and averaged over
all $x$ and $y$ and for $|z| < 2H$, versus box size.  As the domain size is increased, the average temporal variability decreases.}
\label{pv_domain}
\end{figure}

To better quantify this behavior, we define a diagnostic, which
represents the fractional power in temporal fluctuations,
\begin{equation}
\label{pvar}
\pv \equiv \frac{{\rm Var}[\langle f\rangle]}{{\rm Avg}[\langle f\rangle]^2}.
\end{equation}
\noindent
Here, ``Var" refers to the temporal variance of the quantity of interest, and ``Avg'' is the temporal average of the quantity.  Figure~\ref{pv_domain} shows this quantity versus box size for both the total stress (squares) and the magnetic energy (asterisks), calculated from orbit $50$ onward within the ``turbulent" disc ($|z|<2H$). The data of this figure demonstrate that there is a clear decrease in the variability with increasing domain size. Specifically, as we increase the domain size from $2H\times4H\times8H$ ($\two$) to $8H\times16H\times8H$ ($\eight$) (a factor $16$ increase in volume), we find that the fractional variability decreases by approximately a factor $10$. Increasing the volume by a further factor of four (i.e. from simulation $\eight$ to $\sixteeny$) results in only a (comparatively) small decrease in fractional variability, suggesting that we are approaching convergence at this largest domain size.

\begin{figure*}
\begin{center}
\leavevmode
\includegraphics[width=\textwidth]{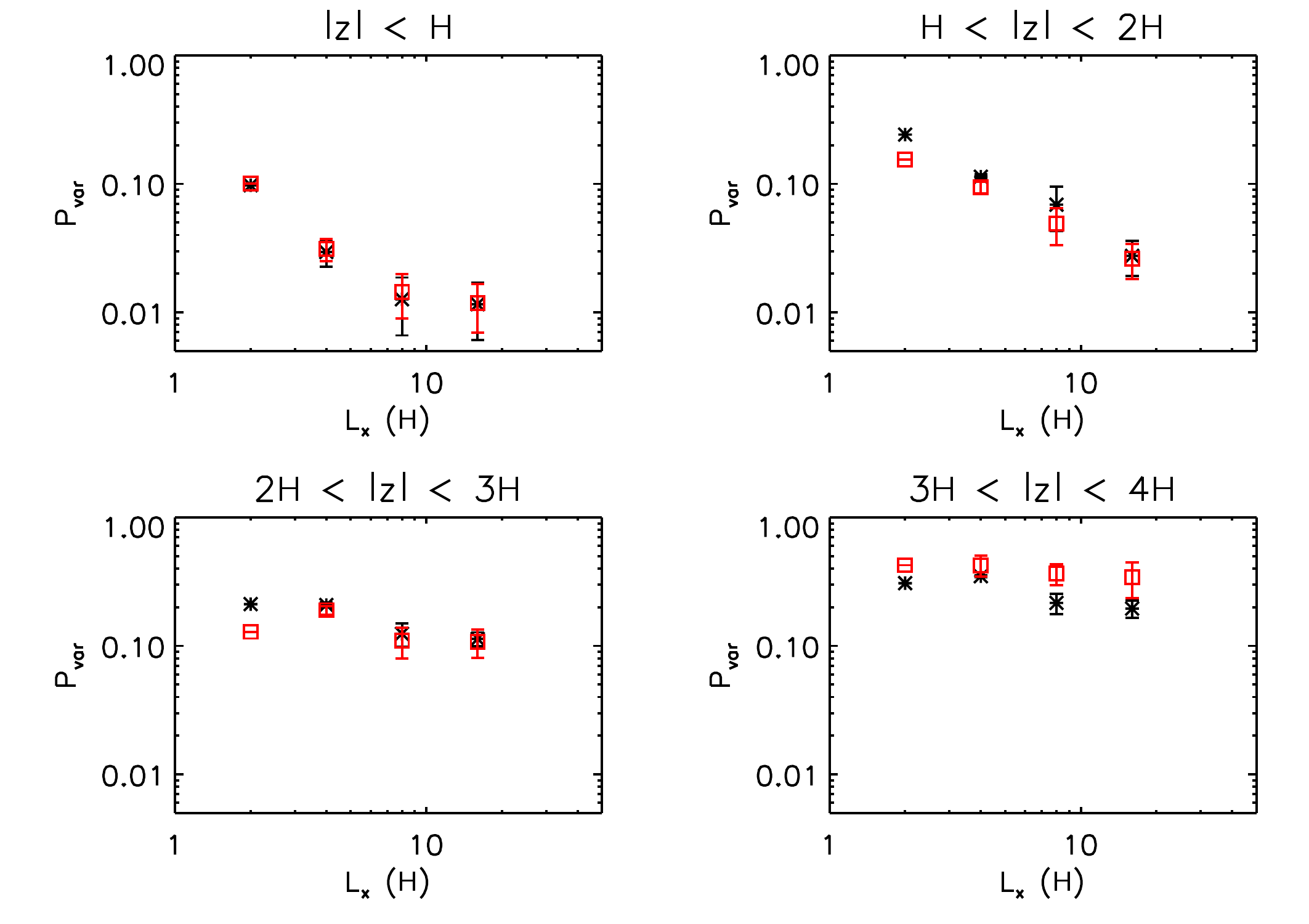}
\end{center}
\caption[]{Variability of the stress (red squares) and magnetic energy (black asterisks) as defined by $\pv$ (equation 12) versus box size.  This is the same quantity as is plotted in Fig.~\ref{pv_domain}.  However, the total domain is broken into horizontal subdomains of size $2H \times 4H$, after which $\pv$ is calculated for that subdomain.  All values of $\pv$ are then averaged together and plotted against the box size.  The error bars denote one standard deviation about the average over subdomains. As labelled, each panel corresponds to a different vertical region in the shearing box.  The primary result of this figure is that near the mid-plane, temporal variability decreases uniformly throughout the grid as domain size is increased, whereas for $|z| \gtrsim 2H$, this variability is constant with domain size.}
\label{var_sd}
\end{figure*}

One might expect that the volume averaged variability will decrease as domain size is increased; as we average over increasing numbers of (largely) uncorrelated  subvolumes, the fractional variability of the system will decrease. If this statistical effect is solely responsible for the observed decrease in variability, we would expect the variability associated with a small domain within the simulation volume to remain unchanged as the volume is increased. To test this notion, we divide each shearing box into horizontal subdomains of size $L_x\times Ly = 2H\times4H$ (i.e. the domain size of simulation $\two$~on the $x-y$ plane), calculate $\pv$ in each subdomain, and then average the $\pv$.  This quantity is plotted in Fig.~\ref{var_sd} for a variety of vertical regions in each shearing box, excluding the smallest, $\half$.  The error bars denote one standard deviation about the average $\pv$ value.  Thus, the error bars do not represent temporal variability but instead quantifies the spread in $\pv$ values between the subdomains.

In the corona $|z| \gtrsim 2H$, the variability of the subdomains remains constant as domain size increases, but in the turbulent disc ($|z| \lesssim 2H$), the variability of the {\em subdomains} decreases with increasing domain size.  As the shearing boxes are increased in size, the {\em physics} of the turbulence changes such that variability within the turbulent disc ($|z| \lesssim 2H$) decreases everywhere uniformly, while the variability in the coronal region remains constant.

%% file: comparison.tex

\section{Comparing the Mesoscale and Global Regimes}\label{compare}

Measured in units of $H$, our largest shearing boxes span radial extents that are not so different from those 
modeled in some global calculations. It is therefore possible to directly compare some of our results 
to those obtained from studies of disc turbulence in the global regime.  We restrict our comparison to recent global simulations 
which attain comparable resolutions to the $32$ zones per $H$ of the local runs 
\cite[such as][]{Beckwith:2011,Sorathia:2011}. The most secure comparisons are probably with the simulations of 
 \cite{Sorathia:2011}; these have orbital advection implemented, which leads
to a significant reduction in numerical diffusion. We note that quantitative comparisons are not 
always possible: there are significant and unavoidable differences in the setup and analysis between the local and global 
runs.  We also note that the scale height defined in \cite{Beckwith:2011} and \cite{Flock:2011b} is a factor of $\sqrt{2}$ smaller than the scale
height defined here.  While this is an order unity effect, it should be borne in mind when making direct comparisons between
that work and the work presented here.

To start, we consider the standard $\alpha$ and $\alpha_{\rm mag}$ parameters.
In all of our calculations, we find that $\alpha \sim 0.02-0.03$, not significantly different than the same parameter calculated (albeit
calculated differently) in global simulations.  Figure 5 of \cite{Beckwith:2011} displays $\alpha$ versus time. While there is a long term decrease
in the $\alpha$ value, their values range from 0.02 to 0.06.  The results of \cite{Flock:2011b} give $\alpha \sim 0.005-0.02$, somewhat
lower than our values and the \cite{Beckwith:2011} values.  \cite{Sorathia:2011} calculated $\alpha$ for various initial field geometries;
they find that $\alpha$ ranges from $\sim 0.01$ to $\sim 0.08$ depending on the field geometry and resolution.

Again considering Fig. 5 of \cite{Beckwith:2011}, the dashed lines in the bottom panel show $\alpha_{\rm mag}$ as a function 
of time, which is relatively constant after the initial growth of the MRI.  The values of $\alpha_{\rm mag}$ range between 0.3 and 0.4, generally consistent
with all of our runs (except for $\half$ as elaborated upon above), for which $\alpha_{\rm mag} \sim 0.4$.  The highest resolved global simulations in \cite{Hawley:2011} have
$\alpha_{\rm mag} \sim 0.2-0.4$, again generally consistent with our shearing box results (though, somewhat lower).  This quantity can be
related to the tilt angle of the magnetic correlation function \citep{Guan:2009a,Sorathia:2011}, which is what \cite{Sorathia:2011} focused on calculating. 
Our typical value of $\alpha_{\rm mag} \sim 0.4$ corresponds to $\theta_{\rm tilt} \sim 12^{\circ}$ in general agreement with the \cite{Sorathia:2011} calculations.  The global calculations of \cite{Flock:2011b} return $\theta_{\rm tilt} \sim 8-9^{\circ}$, slightly
lower than our value as well as that of \cite{Sorathia:2011}.

\begin{figure}
\begin{center}
\leavevmode
\includegraphics[width=\columnwidth]{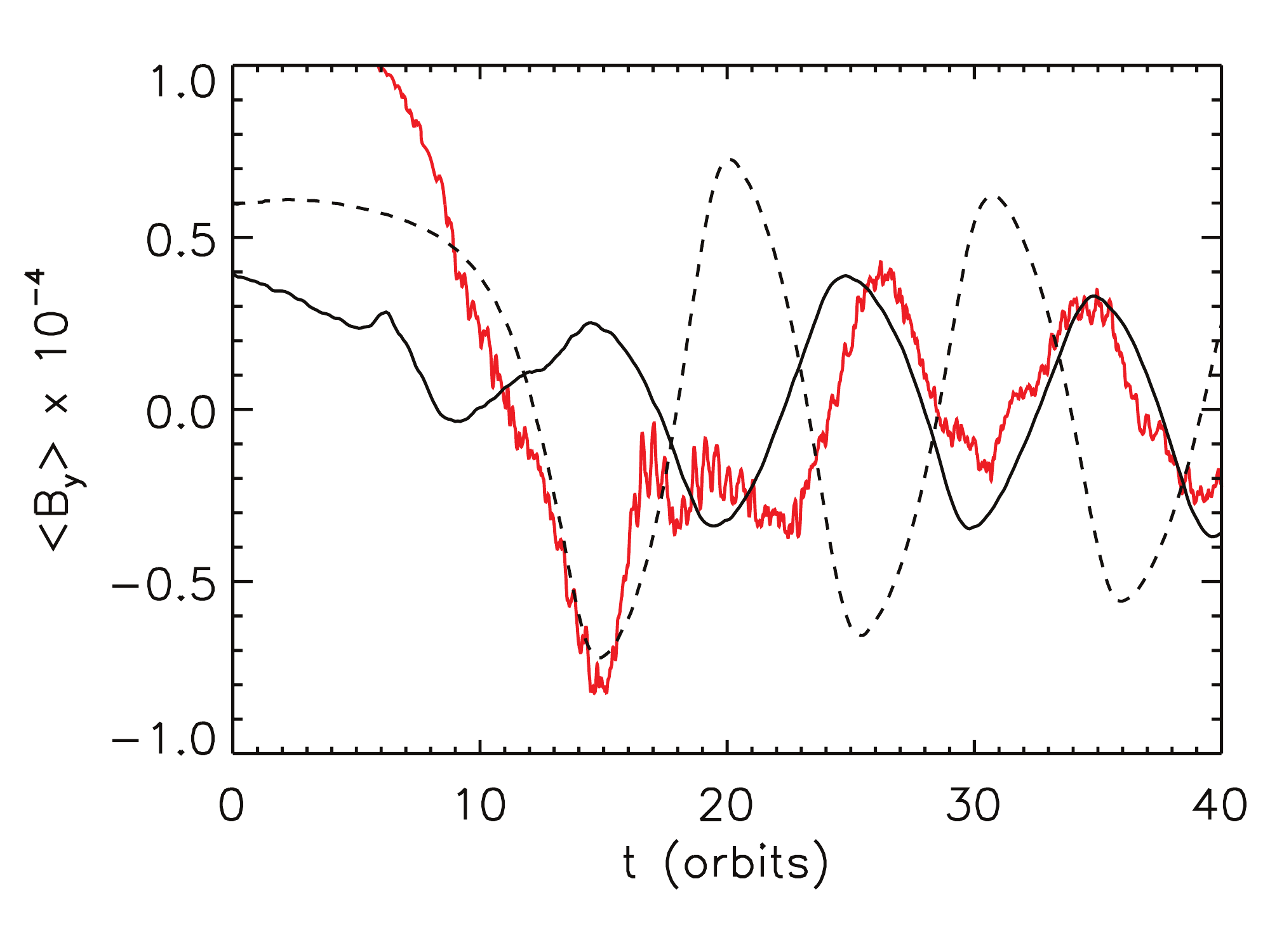}
\end{center}
\caption[]{Time evolution of the volume averaged toroidal field in code units (for all $x,y$ and for $|z| \le 2H$). The solid black curve is $\eight$, the dashed curve is $\sixteeny$, and
the red curve is global simulation data from \cite{Beckwith:2011} (their Fig. 3), renormalized to match the vertical scale of the shearing box data and plotted against the time unit of the local simulations. The early evolution of the net toroidal flux is comparable between the global and local simulations; the initial net flux is rapidly expelled after which the flux oscillates about zero. The original global simulation data from \cite{Beckwith:2011} is for $20$ orbits at $15$ Schwarzschild radii. The agreement between the curves in the figure suggest that the dynamo present in the simulation of \cite{Beckwith:2011} has a characteristic radius of $10$ Schwarzschild radii, at which the duration of each cycle of the dynamo is $10$ orbits and agrees with local simulations.}
\label{flux_global_compare}
\end{figure}

\begin{figure}
\begin{center}
\leavevmode
\includegraphics[width=\columnwidth]{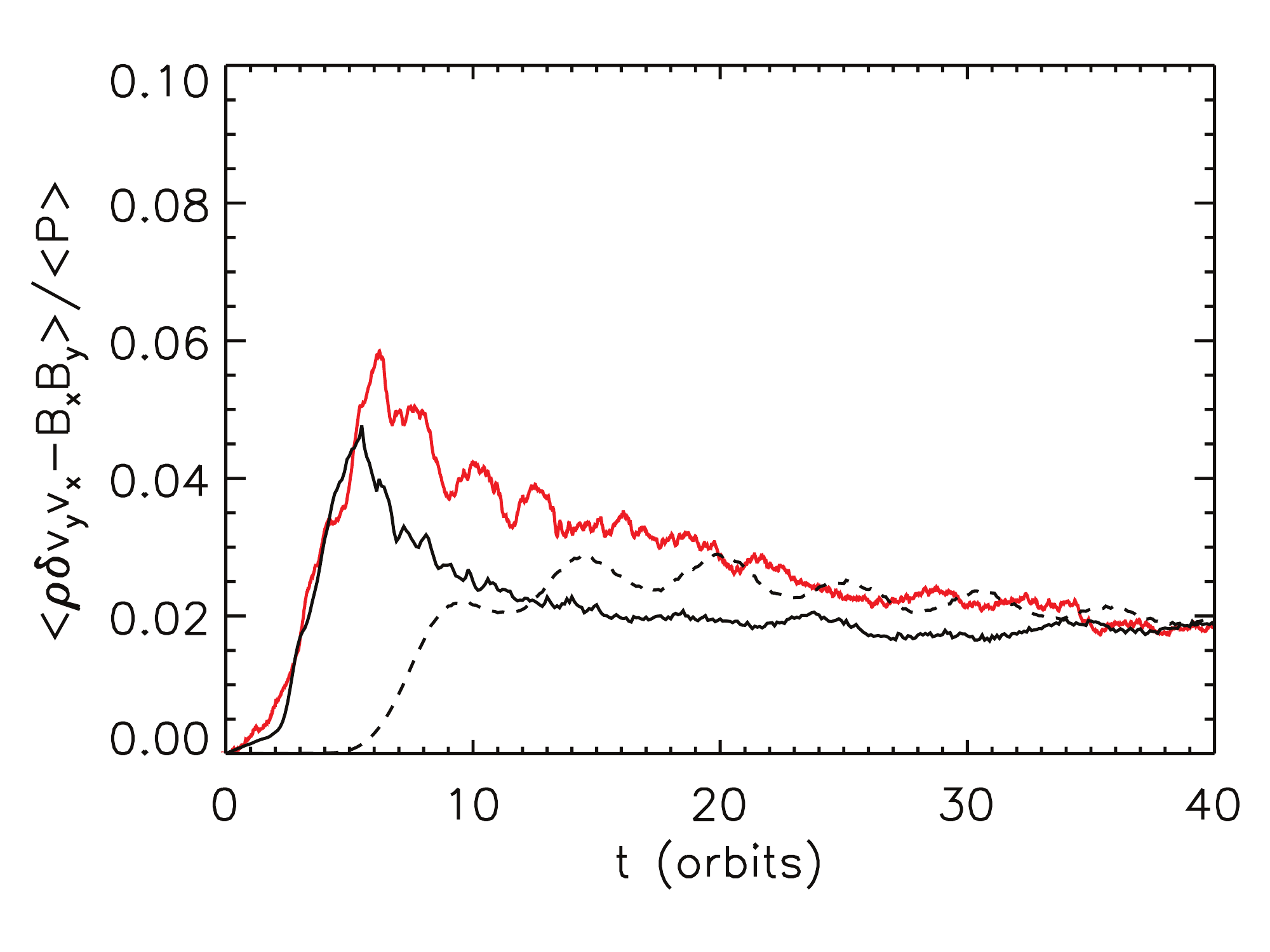}
\end{center}
\caption[]{Volume averaged stress normalized by the volume averaged gas pressure versus time for $\eight$ (black solid line), and $\sixteeny$ (dashed line). This average was done over all $x$ and $y$ and for $|z| \le H$.  The red curve is the equivalent normalized stress from \cite{Beckwith:2011}. The horizontal scale is chosen to match that of Fig.~\ref{flux_global_compare}, and as such, the global data is more representative of a radius of $10$ Schwarzschild radii, as explained in the text and Fig.~\ref{flux_global_compare}.  Similar variability levels are observed in both local simulations and global simulations.  }
\label{stress_1H_global_compare}
\end{figure}

Other behaviors are best compared in a more qualitative manner.   The ``butterfly diagram" appears
to be a robust feature of vertically stratified MRI-driven turbulence.  It has been seen in several global calculations \cite[e.g.,][]{ONeill:2011,Beckwith:2011,Flock:2011a,Flock:2011b},
and the period of the azimuthal field sign flipping is approximately 10 local orbits with power shared radially \citep{ONeill:2011}.  Looking again at the time histories of the pressure-normalized total stress in global simulations, the temporal variability in the global simulations appear to be quite small \citep{Beckwith:2011,Flock:2011b}, similar to the largest shearing boxes presented here. A more explicit comparison can be made by examining the history of the largest shearing boxes presented here over the duration of a well-resolved global simulation. To do so, we note that Fig. 3 of \cite{Beckwith:2011} plots the time-history of the toroidal magnetic flux within the simulation domain, which exhibits transient behaviour, followed by two complete dynamo cycles. In Fig. \ref{flux_global_compare}, we combine this global simulation data with the history of simulation $\eight$ and $\sixteeny$. Note that the local data is plotted over $40$ orbital periods, whereas the data presented by \cite{Beckwith:2011} is for $20$ orbits at $15$ Schwarzschild radii. This would suggest that the dynamo present in the simulation presented in \cite{Beckwith:2011} has a characteristic radius of $10$ Schwarzschild radii, at which the duration of each cycle of the dynamo is $10$ orbits. The figure shows a comparable evolution between the local and global simulations; transient behavior (corresponding to expulsion of toroidal flux from the simulation domain) lasts for $\sim15$ orbits, after which the MRI dynamo produces toroidal magnetic fields of alternating sign on a $10$ orbit timescale. Figure ~\ref{stress_1H_global_compare} shows the stress evolution comparison over this same time period for $\eight$ and $\sixteeny$ averaged over all $x$ and $y$ and for $|z| \le H$. A more direct comparison will be necessary in the future (e.g., calculating the $\pv$ parameter in a shearing box sized region of a global simulation), but this basic comparison suggests that our largest boxes are reaching the
temporal variability level seen in global simulations.
\vspace{0.1in}

The vertical structure of the turbulence, namely the magnetic field, is another point of comparison.  Our calculations show a roughly flat
distribution of the gas pressure normalized stress at $\sim 0.01$ for $|z| \lesssim 2H$, after which the stress drops to $\sim 10^{-4}$ (see Fig.~\ref{all_z}).  The
same figure shows that $\beta$ peaks near 40 and then drops below unity outside of $|z| = 2H$.  In \cite{Beckwith:2011}, the normalized stress peaks near $\sim 0.01$ and
$\beta \sim 30$ at its peak, generally consistent with our results.  However, the behavior away from the mid-plane differs dramatically from what is seen in our local
calculations.  In particular, the stress does not drop nearly as rapidly outside of $|z| = 2H$, and $\beta$ never drops below unity.  This difference in behaviors
away from the mid-plane could be due to the use of periodic boundary conditions in the global simulations of \cite{Beckwith:2011}.  The general structure of
$\beta$ in the global BO model of \cite{Flock:2011a}, which has vertical outflow boundary conditions, (their Fig. 11) is similar to what we observe in our local calculations except that everything appears to be scaled
up by a factor of $\sim 10$ in \cite{Flock:2011a}.  That is, their peak $\beta$ value is $\sim 500$, and at $|z| \approx 2H$, $\beta \sim 10$.  The reason for this
large difference is not at all clear. 

All of these comparisons, with the exception of the vertical magnetic structure, yield good agreement between local and global calculations, considering
the spread in resolution and measurement techniques.  However, the real question we have set out to answer in this paper is whether or not the mesoscale appears the same between these two calculations; this has yet to be addressed.  The most relevant measure for answering this question is the correlation function of various turbulent structures in the horizontal plane.  

The analysis of \cite{Beckwith:2011} and \cite{Flock:2011b}, shows a tilted, highly localized component to turbulent fluctuations, in agreement with the results found in this work.  In their analysis, \cite{Beckwith:2011} subtract off large scale variations (due to radial and vertical gradients) to only focus on the central, tilted component.   They then calculate a correlation length along the major and minor axes of the tilted structure by determining the length at which the correlation function drops by a factor of ${\rm e}^{-2}$ (see their Table 1).  \cite{Flock:2011b} calculates the same lengths but by using the half width at half maximum of the correlation function.    The correlation length values between these two papers generally agree, though it should be noted that \cite{Beckwith:2011} consider the ACF of the magnetic energy, whereas \cite{Flock:2011b} calculate
the ACF for ${\bm B}$ as was done here and in \cite{Guan:2009a}.

In our calculation of the ACF, we do not subtract off any background component.  However, we can consider the correlation length along a given axis to be the point at which the ACF begins to flatten out.  Beyond this length, the ACF is flat because the structure is relatively uniform; thus, the "knee" in the curve represents the outermost scale of the localized component to the ACF.  From Fig.~\ref{corr1d}, it appears that the correlation lengths for ${\bm B}$ are $\lambda_{B,{\rm maj}} \sim 4-5 H$ and $\lambda_{B,{\rm min}} \sim 0.3-0.4 H$.  Viewing the equivalent plot for $B^2$ and $\rho$ gives $\lambda_{B^2,{\rm maj}} \sim 2-3 H$, $\lambda_{B^2,{\rm min}} \sim 0.2 H$, and $\lambda_{\rho,{\rm min}} \sim 0.3-0.4 H$.  We did not calculate a major axis correlation length for $\rho$ because the ACF never flattened out along the major axis.  This is likely related to the presence of the axisymmetric zonal flows entering into the ACF calculation, though it is still not clear why the ACF does not flatten out at some point along the major axis.  
In any case, it would seem that there is rough agreement between the correlation length of the tilted, localized ACF component in global and local calculations.

The general agreement between these correlation lengths is very encouraging as it shows that MRI-driven turbulence has the same small scale structure in both local and global simulations.  However, a primary result of the work here is that in addition to the tilted component of the magnetic field structure, there is also a non-negligible, volume filling background component.  Again, since the large scale component was subtracted off in the analysis of \cite{Beckwith:2011}, a direct comparison with our results cannot be made very easily.  However, some information can be extracted by examining their Fig. 4, which shows the structure of the toroidal magnetic field on the azimuthal plane. The figure shows that this field has structure in both the vertical and radial dimensions; there appear to be magnetic field bundles that evolve throughout the disc. Through a very rough examination of this figure, it appears that the radial size of these bundles is $\sim 5 r_s$, where $r_s$ is the Schwarzschild radius.  In those calculations, $H/R = 0.07$ so that at $R = 10 r_s$, $H = 0.7 r_s$.  Thus, the typical size of the magnetic field bundles in that figure is estimated to be $\sim 7H$ where (as noted above) this $H$ is that defined in \cite{Beckwith:2011}, a factor of $\sqrt{2}$ smaller than the $H$ that we use.  This scale translates to $\sim 5H$ using $H$ defined as we have in this paper, which should definitely be captured in our largest shearing box, $\sixteeny$.  However, the correlation function of ${\bm B}$ clearly indicates that there is a volume-filling background component to the magnetic field; there is no structure associated with $\sim 5H$.  Granted, this comparison is murky at best, and perhaps even larger shearing boxes would show the presence of such radial structure.  However, these results suggest that there may be a fundamental difference in radial magnetic structure at the mesoscale between local and global simulations, perhaps a result of the inclusion of curvature terms in global simulations.

The presence of density zonal flows in global calculations done to-date is questionable at best.   To our knowledge, the best evidence for such flows exist in the 
global calculations of \cite{Lyra:2008}; see their Fig. 10, which shows reasonably long-lived radial variations in the gas density.  However, even here, the flows are only weakly detected and not discussed by the authors in any detail.  It would thus seem that these flows are very difficult to detect in global calculations.  This difficulty may result from larger scale density gradients dominating the weaker zonal flow variations.   While our work here has shown tentative evidence for a converged outer scale to these flows in shearing boxes, it is still not clear whether or not such flows are an artifact of the shearing box approximation that tend to exist on the largest radial scales.  Thus, we believe that the detection of these structures in global simulations should be a very important priority for future work, one that we are currently pursuing.

%% file: conclusion.tex
\section{Discussion, Summary, and Conclusion}
\label{discussion}

\begin{figure}
\begin{center}
\leavevmode
\includegraphics[width=\columnwidth]{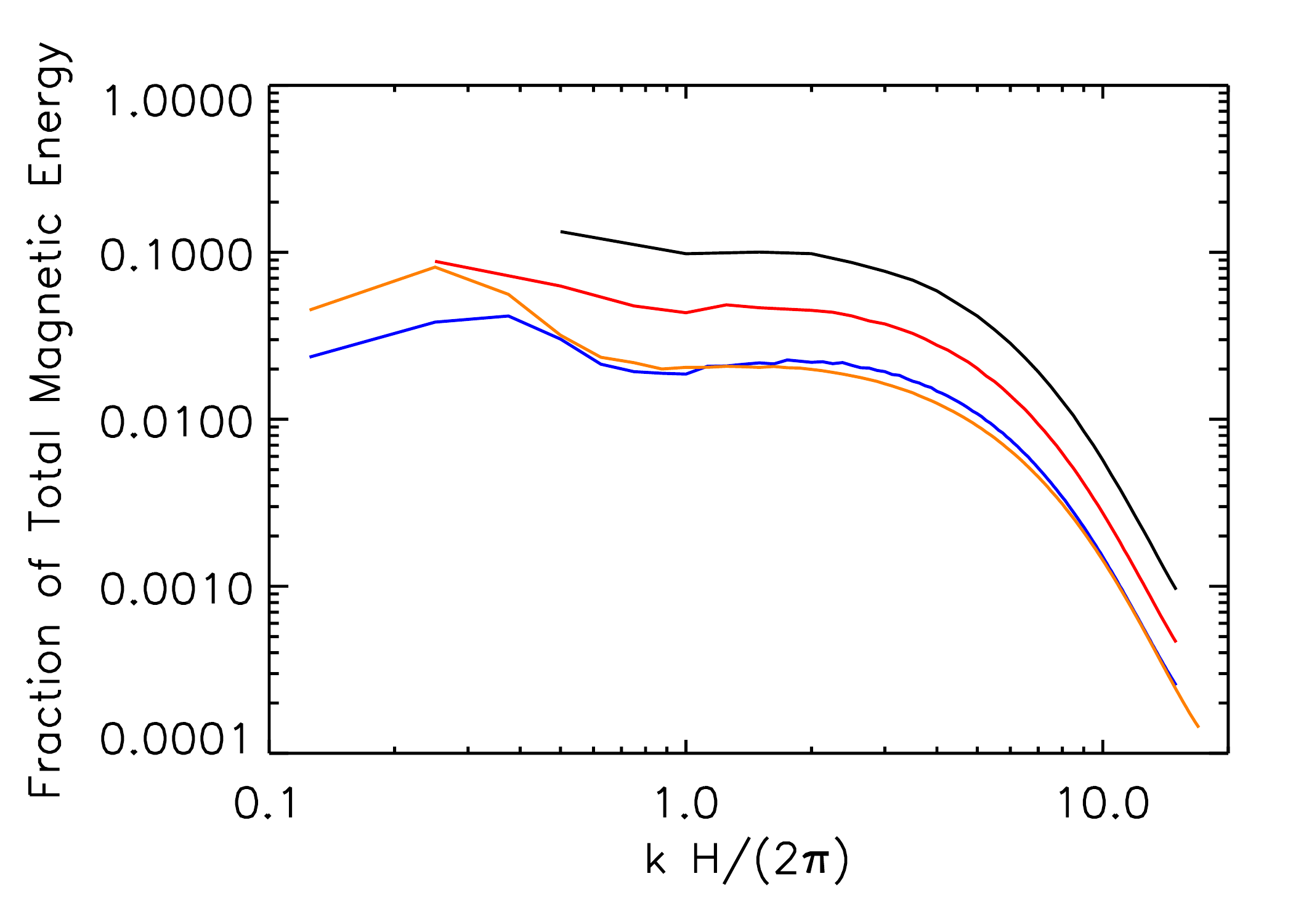}
\end{center}
\caption[]{Fraction of the total magnetic energy as a function of isotropic wavenumber $k$.  In calculating these curves, we averaged the three-dimensional
Fourier transform of magnetic energy over spherical shells of constant $k$ and then averaged the resulting one-dimensional power spectra in time, from orbit 50 to 125.
We omit data from the smallest domain (\half) for the purposes of clarity. The black curve is from $\two$, red is $\four$, blue is $\eight$, and orange is $\sixteeny$.   As domain
size is increased, the fraction of the total magnetic energy at the smallest scales decreases while the largest fraction of magnetic energy resides at the largest scales.  There appears to be convergence between the two largest domain sizes.}
\label{ps_be}
\end{figure}

We have examined the convergence properties of MRI-driven MHD turbulence with increasing domain size, i.e. as we transition from the ``local'' ($H \sim L \ll R$) to ``mesoscale'' ($H \ll L \ll R$) regime. We have found that volume-averaged quantities \cite[such as the][$\alpha$-parameter and the ratio of the Maxwell stress to magnetic energy, $\alpha_{\rm mag}$]{Shakura:1973} converge rapidly as we transition between these regimes. We have also studied the properties of the magnetic field autocorrelation function (ACF) as we move between these regimes, which reveals a two component structure to the magnetic field.  One is highly localized and
tilted with respect to the azimuthal direction, consistent with previous findings \cite[e.g.,][]{Guan:2009a}.  The other component is
extended to the largest scales in the domain, and is likely associated with the (predominantly toroidal) background magnetic field. Furthermore, we have identified the zonal flows of \cite{Johansen:2009} in our domain sizes of scale $4H\times8H\times8H$ and larger. These flows always fill the largest radial scale in the box except for our largest simulation.  We find tentative evidence that the radial wavelength of these flows converge at $\sim 12H$.

These results have important implications for the physics of angular momentum transport in magnetized accretion discs. While the ACFs of the Maxwell and Reynolds stress are strongly concentrated within $\lesssim H$, both of these quantities are also correlated on scales $\gg H$ at the $\sim20\%$ and $\sim8\%$ level respectively. If angular momentum transport in magnetized accretion discs is a purely local process, then the ACF of these quantities should drop to zero on scales $\sim H$ \citep{Gammie:1998}. That the ACFs of the accretion stress remain finite at scale $\gg H$ undercuts the assumption that angular momentum transport in magnetized discs can be treated as a purely local process \citep{Shakura:1973,Pringle:1981}. 

We have also found that the transition to the mesoscale limit is associated with a decrease in the temporal variability of the turbulent disc, an effect that is not solely due to volume-averaging over uncorrelated subdomains. It is perhaps not surprising that the small scale physics of MRI-driven turbulence changes with increasing box size; as larger magnetic structures are permitted in the box, large
scales can potentially interact with smaller scales and change the variability properties at these scales.  While a spectral energy transfer analysis \citep{Fromang:2007a,Fromang:2007b,Simon:2009a} would be useful in understanding how exactly different scales interact, we can attain a basic understanding by considering the time-averaged power spectra of the magnetic energy, as shown in Fig.~\ref{ps_be}.  In calculating this power spectra, we took a full three-dimensional Fourier transform, averaged
the result over spherical shells of constant $k$, and then time averaged the resulting spectra from orbit 50-125.  The figure shows the fraction of the total magnetic energy that exists at a scale $k$ (i.e., integrating each curve over $k$ would return unity).  As box size is increased, the power spectra shift downwards because there are more spatial scales over which the magnetic energy can be distributed.  However, the main point to take away from this figure is that as the box size is increased, the fraction of the total magnetic energy at the smallest scales generally decreases while the largest fraction of magnetic energy resides at the largest scales, with signs of convergence between the two largest domain sizes, \cite[this result is consistent with Fig. 2 of][]{Johansen:2009}. Since the turbulent energy levels are roughly the same between all of these simulations (e.g., Fig.~\ref{turb_avg}), increasing the domain size equates to more power being taken from the small scales and put into the largest scales.  How and why exactly this behavior occurs is still an open question.  However, this result is consistent with our notion that the largest scales in the shearing box play a very important role in the evolution of
the MRI.

We have also examined the physical properties of shearing boxes computed using restricted spatial dimensions, e.g. $0.5H\times2H\times8H$ in $L_x\times L_y \times L_z$. Turbulence arising from the MRI in these simulations is characterized by a greater level of isotropy ($\alpha_{\rm mag} = 0.16$) than is found in larger shearing box domains (those with $L_x \ge 2H \times L_y \ge 4H$). Furthermore, we find anomalous structures in both the ACFs and vertical profiles of (e.g.) the magnetic energy. These results suggest that simulation domains of this size or smaller could produce misleading results.  One such use of small domains (necessary for numerical reasons) is in
radiation pressure dominated MRI simulations \cite[e.g.,][]{Hirose:2009,Blaes:2011}.  However, the ACF of the magnetic field in these simulations strongly resembles
that of our converged $\two$ calculation (Hirose, private communication), despite the fact that the scale height of these radiation pressure dominated simulations is not
very different from an isothermal disk scale height as we have in our calculations (Krolik, private communication).  This surprising result indicates that the thermodynamic properties of the disk may play a role in determining the detailed structure of MRI-driven turbulence. 

We have compared results from the largest domain sizes considered here ($8H\times16H\times8H$ and $16H\times32H\times8H$) with those obtained from well-resolved global simulations that have been recently presented in the literature \cite[e.g.,][]{Beckwith:2011,Sorathia:2011}. We have found broad agreement between these two classes of simulations, in terms of volume-averaged quantities, the vertical structure of and correlations lengths within the turbulence. A point of contention however, is the largest spatial scale on which (e.g.) the magnetic field is correlated. The mesoscale simulations presented here suggest that correlations exist on radial scales up to the maximum considered ($16H$). Crude analysis of the results of \cite{Beckwith:2011} suggests, however, a maximum radial correlation length of $\sim5H$. This result may indicate that the maximum radial correlation length of the magnetic field is determined by global (e.g., curvature), rather than local effects. Furthermore, if mesoscale structures in the magnetic field and accretion stress originate in the action of the toroidal field MRI, then global simulations that probe up to $m=1$ in azimuth, utilizing resolutions equivalent to the simulations presented here, will be necessary to determine the outer scale of these quantities. Since there are non-negligible levels of angular momentum transport at the largest azimuthal scales, a first-principles global treatment of the system is necessary to understand the complete physics of angular momentum transport.

Overall, our results suggest several avenues of future work. If, as suggested by our results and those of \cite{Beckwith:2011}, a significant amount of angular momentum transport takes place on scales $\gg H$, it will be important to determine the length scale on which the work done by this large scale angular momentum transport is dissipated. \cite{Hirose:2011} suggest that the magnetic fields associated with this large scale angular momentum transport can rise buoyantly into the disc corona ($|z|>2H$), where they can be dissipated. If this is the case, then the local connection between angular momentum transport and disc heating would be broken, further undermining the assumptions of local models of angular momentum transport \citep{Balbus:1999}. It would then be important to understand the mechanisms by which the dissipated heat would be radiated away by the disc and any resulting changes to the vertical structure. 

Equally important will be to identify the physical mechanism that causes the decrease in turbulent variability as we transition to the mesoscale regime. As it currently stands, our work suggests that the variability of an ionized accretion disc is tied to the global structure of the flow. These results, in combination with our results regarding the locality of angular momentum transport, suggests that attempts to interpret observations of accretion variability through local, viscous models of angular momentum transport \citep{King:2007,Ingram:2011} could yield misleading conclusions regarding the \emph{physics} of magnetized accretion discs. Future global simulations that incorporate boundary conditions and physics appropriate to a given astrophysical system are likely to be necessary to provide accurate interpretations regarding the variability of magnetized accretion discs.

Finally, the detection of density zonal flows in global calculations will be a very important avenue to pursue.  As mentioned above, it will be necessary to determine whether these flows are an artifact of the shearing box approximation or are in fact present in global simulations as well.  If they do exist in global calculations, we will need to quantify their radial (and azimuthal) scale as well as their amplitude relative to the background density.  Such quantifications will be important in informing models of planetesimal trapping and planet migration.